# Accountability Infrastructure:
# How to implement limits on platform optimization to protect population health

June 2023

Nathaniel Lubin
Thomas Krendl Gilbert

# EXECUTIVE SUMMARY

Attention capitalism has generated design processes and product development decisions that prioritize platform growth over all other considerations. To the extent limits have been placed on these incentives, interventions have primarily taken the form of content moderation. While moderation is important for what we call "acute harms," societal-scale harms – such as negative effects on mental health and social trust – require new forms of institutional transparency and scientific investigation, which we group under the term accountability infrastructure.

This is not a new problem. In fact, there are many conceptual lessons and implementation approaches for accountability infrastructure within the history of public health. After reviewing these insights, we reinterpret the societal harms generated by technology platforms through reference to public health. To that end, we present a novel mechanism design framework and practical measurement methods for that framework. The proposed approach is iterative and built into the product design process, and is applicable for both internally-motivated (i.e. self regulation by companies) and externally-motivated (i.e. government regulation) interventions for a range of societal problems, including mental health.

We aim to help shape a research agenda of principles for the design of mechanisms around problem areas on which there is broad consensus and a firm base of support. We offer constructive examples and discussion of potential implementation methods related to these topics, as well as several new data illustrations for potential effects of exposure to online content.

# TABLE OF CONTENTS

# GLOSSARY

**Platform** – A digital interface through which users can generate, upload, and share content at scale.

**Population** – A particular group of people who are similarly situated geographically or via other defined attributes, such as shared demographics, and are similarly influenced by platforms.

**Acute harms** – Injury or damage inflicted over a short period of time via the discrete action of one individual (or system) on another individual.

**Structural harms** – Injury or damage inflicted over time via exposure to a series of damaging objects or systems, producing adverse effects on a defined population.

**Public health** – The science of protecting and improving the health of people and their communities.

**Mechanism design** – A set of rules imposed on actors within a system that induces a set of actions or behaviors by those actors in order to achieve a defined objective.

**Accountability infrastructure** – A system capable of assessing and mitigating societal harms – both acute and structural – on at-risk populations. A fully-actualized version will include systems for selecting which harms to address and which metrics to leverage for assessment.

**Content review / content moderation** – The practice of individualized review and potential removal of digital material posted by third-party users, generally predicated upon a written set of rules (e.g. terms of service or codes of conduct).

**Attention Economy** – Systematized revenue-seeking activity based on maximizing the time spent by users, typically via advertising.

**Controllable interventions** – Experiments intended to inform product design changes based upon statistically-valid randomized controlled trials.

*This is an era of specialists, each of whom sees his own problem and is unaware of or intolerant of the larger frame into which it fits. It is also an era dominated by industry, in which the right to make a dollar at whatever cost is seldom challenged. When the public protests, confronted with some obvious evidence of damaging results of [pesticide] applications, it is fed little tranquillizing pills of half truth. We urgently need an end to these false assurances, to the sugar coating of unpalatable facts. It is the public that is being asked to assume the risks that the insect controllers calculate. The public must decide whether it wishes to continue on the present road, and it can do so only when in full possession of the facts. In the words of Jean Rostrand, "The obligation to endure gives us the right to know."*

– Rachel Carson, *Silent Spring* (1962)[1]

# INTRODUCTION

The need for new ways to address the effects of abusive content on tech **platforms** has become a dominant thread across media, technology, financial, and political circles. Responses to these challenges have contributed to events as wide-ranging as Elon Musk's purchase of Twitter,[2] legal responsibilities by actors participating in the January 6 insurrection,[3] and the effectiveness of the COVID-19 response.[4] Meanwhile, Florida and Texas have passed laws designed to roll back supposedly excessive moderation decisions made by social media platforms;[5] and in May 2023, Montana became the first state to officially ban Tiktok.[6] Beyond its recent rulings pushing off larger adjudication of technology moderation (Twitter Inc. v Taamneh and Gonzalez v. Google Inc.), it remains unclear whether and how the Supreme Court will engage with Section 230 of the Communications Decency Act.[7] In summary, digital media law, mostly unchanged since the 1990s, is undergoing a tectonic shift.

---

[1] Carson, Rachel. *Silent Spring*. Penguin, 2002. p. 13.
[2] Zakrzewski et al. "Musk's 'free speech' agenda dismantles safety work at Twitter, insiders say." *The Washington Post*, 22 Nov. 2022.
https://www.washingtonpost.com/technology/2022/11/22/elon-musk-twitter-content-moderations/
[3] Lapowsky, Issie. "Jan. 6 launched a wave of anti-content moderation bills in America." *Protocol,* 6 Jan. 2022.
https://www.protocol.com/bulletins/anti-content-moderation-bills
[4] Ferreira Caceres, Maria Mercedes et al. "The impact of misinformation on the COVID-19 pandemic." AIMS public health vol. 9,2 262-277. 12 Jan. 2022, doi:10.3934/publichealth.2022018
[5] Brannon, Valerie. "Free Speech Challenges to Florida and Texas Social Media Laws." *Congressional Research Service.* 22 Sep. 2022. https://crsreports.congress.gov/product/pdf/LSB/LSB10748
[6] Shepardson, David. "Montana to become first US state to ban TikTok." *Reuters,* 18 May 2023.
https://www.reuters.com/world/us/montana-governor-signs-bill-banning-tiktok-state-2023-05-17/
[7] For example: Anderson, Scott et al. "The Supreme Court Punts on Section 230." *Lawfare,* 19 May 2023.
https://www.lawfareblog.com/supreme-court-punts-section-230

Despite serious engagement by governments on both sides of the Atlantic, including significant laws passed by the EU and UK, easy solutions are not forthcoming.[8] As platforms have become more complex, it has become increasingly clear that content review alone is not a scalable, reliable, or effective leverage point for many regulatory goals.[9] Nor is this problem specific to social media platforms, as illustrated by the controversial rollout of new AI tools like ChatGPT.[10] Evaluation and potential restrictions on new tools will not be effective based on the direct effects of its use (e.g. teachers who fear the submission of AI-written papers[11]). Instead, any serious restrictions will need to engage with the aggregate, emergent effects of the product and how any particular product interacts with discrete, identifiable populations. For instance, the legal engagement by communities of artists whose work was used as training data in models (without compensation or consent) poses a regulatory question, regardless of how courts respond.[12]

Most attempts to engage with these challenges – especially with new proposals for improved measurement – are interpreted either as an ineffectual stopgap that fails to alter the status quo, or as so intractable with respect to existing pipelines that arguing for their implementation comes off as naive. We do not pretend to resolve this binary. On the contrary, we see it as the same story that played out during prior developments of new technology. Just as new rules were needed to meet the public health challenges of the 19th and 20th centuries, platforms require **accountability infrastructure** that places limits on optimization in order to engage at-risk populations safely and legitimately. By accountability infrastructure we aim to capture the range of mechanisms from design interventions to regulatory standards needed to fully evaluate the system-level outcomes of societal scale platforms.[13]

---

[8] Allen, Asha and Ophelie Stockhem. "A Series on the EU Digital Services Act: Ensuring Effective Enforcement." *Center for Democracy & Technology,* 18 Aug. 2022.
https://cdt.org/insights/a-series-on-the-eu-digital-services-act-ensuring-effective-enforcement/
[9] See: Perrigo, Billy. "Inside Facebook's African Sweatshop." *Time,* 17 Feb. 2022.
https://time.com/6147458/facebook-africa-content-moderation-employee-treatment/ or Newton, Casey. "The Trauma Floor, The secret lives of Facebook moderators in America." *The Verge,* 25 Feb. 2019.
https://www.theverge.com/2019/2/25/18229714/cognizant-facebook-content-moderator-interviews-trauma-working-conditions-arizona
[10] For example: Pogue, David. "AI experts on whether you should be "terrified" of ChatGPT." *CBS News,* 22 Jan. 2023. https://www.cbsnews.com/news/ai-experts-on-chatgpt-artificial-intelligence-writing-program/
[11] See: Herman, Daniel. "The End of High-School English." *The Atlantic,* 9 Dec. 2022.
https://www.theatlantic.com/technology/archive/2022/12/openai-chatgpt-writing-high-school-english-essay/672412/
or Huang, Kalley. "Alarmed by A.I. Chatbots, Universities Start Revamping How They Teach." *The New York Times,* 16 Jan. 2023.
https://www.nytimes.com/2023/01/16/technology/chatgpt-artificial-intelligence-universities.html
[12] Indeed, this case offers an interesting question with the questions posed by Section 230 litigation. See: Vincent, James. "The lawsuit that could rewrite the rules of AI copyright." *The Verge,* 8 Nov. 2022.
https://www.theverge.com/2022/11/8/23446821/microsoft-openai-github-copilot-class-action-lawsuit-ai-copyright-violation-training-data
[13] We are aware of forthcoming work from Elettra Bietti ("From Data to Attention Infrastructures: Regulating Extraction in the Attention Platform Economy") that captures some of the dimensions of our proposal, under the label "attention infrastructure." Our proposal may be seen as a way to lend accountability to attention infrastructure at distinct levels of abstraction (in particular technical, organizational, legal, and regulatory).

In general, the complexity of platforms obfuscates the effects of important decisions, including via management processes that incorporate choices made by many different actors.[14] A robust form of engagement must ask questions that are relevant, answerable, and matched to metrics that may be improved through iterative methods.

There are important historical parallels to the present moment. During industrialization in the 19th century, unconstrained, growth-oriented design undercut public health. As harms became more manifest and persistent, effective methods *were* adopted to address them. The critical step was recognizing that the maturation of effective policy frameworks required the simultaneous development of infrastructure to support new forms of intervention, measurement, and evaluation. Today, as the day-to-day experiences of populations have shifted online, there is a similar opportunity to develop new infrastructure that interfaces between emerging technologies and desirable policies.

Two related points are critical in considering this approach. First, even in advance of robust causal explanations, what is first needed is *the capacity to observe causal connections* in the form of actionable metrics. Second, robust causal explanations must be developed in tandem with *infrastructure for interventions and policy-setting*. Our goal in this paper is to outline both dimensions, identifying clear historical lessons for each and applying these to the context of societal-scale online platforms. While the questions remain open regarding how responsibilities for these capacities might be delineated across agents or institutions, we hope this work helps identify common ground on which constructive debate might proceed.[15]

The present model of attention-maximization for algorithmically-driven technology products is largely unrestricted, despite "growth at all costs" cultures and incentives that are embedded in many if not most companies. We argue that the best way to address systemic harms caused by these platforms is to build direct assessments of population health into product development. Such an approach would set standards for a "do no harm" principle on these metrics, alongside the growth-oriented metrics that product teams prioritize in their uses of "objectives and key results" (OKRs). By iteratively developing and improving these types of systems, randomized controlled trials that effectively assess harms can be incorporated directly into product development cycles. In particular, this should be done beginning with areas for which there is the most pre-existing agreement on the importance of the structural harm, so that the implementation of correctives around these harms would not have an explicitly political valence. Through those lenses, we argue that the initial developments of these regimes could begin with a focus on protections of mental health and of measures of societal trust.

---

[14] Much work in the literature on complex systems, organizational theory, and distributed cognition has drawn attention to this problem. For a classic example, see Hutchins, Edwin. *Cognition in the Wild*. MIT press, 1995.

[15] Delineating the roles that can or should underlie this regime of accountability is an active research topic. We see our work as aligned with the discussion in Moss, Emanuel, et al. "Assembling accountability: algorithmic impact assessment for the public interest." *Available at SSRN 3877437* (2021).

We first explore some of the context and motivations for our approach, distinguishing structural harms associated with attention capitalism from the acute harms that are presently addressed via content moderation. Section 2 contextualizes that dichotomy via the history of public health, focusing especially on the initial rise of infrastructure in the mid 19th century. In Section 3, we interrogate the mechanism design questions posed by social media platforms, illustrating what would be organizationally necessary to add meaningful constraints on product design. Section 4 discusses the relation between these ideas and outstanding questions about how to govern and manage societal-scale digital media platforms, including a brief engagement with free speech questions. Section 5 concludes.

# CONTEXT & MOTIVATIONS

Initially celebrated and more recently decried, social media powered by the **attention economy** has generated consistent scrutiny. Scholars have theorized how these platforms transform engrained norms for privacy, autonomy, dignity, and social identity. While culturally profound, these transformations are downstream of a much more banal feature of current engineering practices: these tools maximize the engagement of users at all costs,[16] and any material shift would result in immediate and tangible effects on each platform's userbase.[17] However, the single-minded focus on attention maximization gives companies little reason to enact such shifts.[18]

The builders and proselytizers of these platforms mainly act as advocates for these practices, viewing growth as good for socialization and inevitable in the pursuit of societal-scale platforms. Information exchange is assumed to have intrinsic social and economic value.[19] However, once platforms scale to include the dynamics of population interactions, attention maximization qualitatively shifts the incentive structure of how content is created, distributed, and consumed. It is not at all clear that the effects of the resulting feedback loops between platforms and populations are normatively good,[20] nor whether our inherited understandings of "good" and "bad" outcomes even apply to systems operating at this scale of reinforcement.

To move beyond the present limitations of the attention economy, platforms must instead adopt what we call **accountability infrastructure** for recommender systems. The key differences between these two paradigms are summarized in Table 1.

---

[16] For example: Montag, Christian et al. "Addictive Features of Social Media/Messenger Platforms and Freemium Games against the Background of Psychological and Economic Theories." International journal of environmental research and public health vol. 16,14 2612. 23 Jul. 2019, doi:10.3390/ijerph16142612 and Andersson, Hilary. "Social media apps are 'deliberately' addictive to users." *BBC News,* 4 Jul. 2018. https://www.bbc.com/news/technology-44640959

[17] Price, Catherine. "Trapped - the secret ways social media is built to be addictive (and what you can do to fight back)." *BBC Science Focus,* 29 Oct. 2018. https://www.sciencefocus.com/future-technology/trapped-the-secret-ways-social-media-is-built-to-be-addictive-and-what-you-can-do-to-fight-back/

[18] For example: Kang, Cecilia and Sheera Frenkel. *An Ugly Truth: Inside Facebook's Battle for Domination.* Harper, 2021.

[19] Swisher, Kara. "Zuckerberg: The Recode Interview." *Vox,* 8 Oct. 2018. https://www.vox.com/2018/7/18/17575156/mark-zuckerberg-interview-facebook-recode-kara-swisher

[20] Philosophers have recently characterized the effects as *noxious*, i.e. toxic to important human values. See Castro, Clinton, and Adam K. Pham. "Is the attention economy noxious?." *Philosophers' Imprint* 20.17 (2020): 1-13.

**Table 1:** Rival paradigms of the present and potential landscape of social media regulation.

| Institutional paradigm | Attention Economy | Accountability Infrastructure |
|---|---|---|
| Goal | User attention | Population health |
| Locus of harms | Acute: output-based | Structural: outcome-based |
| Governance frame | Rules-based | Empirics-based |
| Intervention mechanism | Content moderation | Constraints on system behavior |

Absent compelling societal harms demonstrated empirically, the adoption of accountability infrastructure reduces to the default system under the attention economy. Said another way, companies are able to maximize profits or whatever interests they choose up until the point where population health is meaningfully affected.

## On the Need for Oversight: Reassessing the Attention Economy

The rise of social media platforms inaugurated a shift in content generation from providers to users.[21] We use the term "**platform**" capaciously to refer to web interfaces through which users can generate, upload, and share content at scale. So defined, major platforms include Facebook (2004), YouTube (2005), Reddit (2005), Twitter (2006), Pinterest (2010), Instagram (2010), Discord (2015), and TikTok (2016). They also include more recent interfaces such as the open source networking service Mastodon (2016), APIs for content generation like DALL-E (2021) and ChatGPT (2022), as well as the integration of these technologies into prominent search engines, e.g. Bing (2023).[22] Platforms have an ambiguous technical and political status, both enabling and constraining the ability to share content.[23,24]

---

[21] For example: Grand View Research Inc. "User Generated Content Platform Market Size Worth $18.65 Billion By 2028: Grand View Research, Inc." *Cision PR Newswire*, 19. Jul. 2021.
https://www.prnewswire.com/news-releases/user-generated-content-platform-market-size-worth-18-65-billion-by-2028-grand-view-research-inc--894027798.html
[22] Bhuiyan, Johana. "Microsoft reportedly to add ChatGPT to Bing search engine." *The Guardian,* 5 Jan. 2022. https://www.theguardian.com/technology/2023/jan/05/microsoft-chatgpt-bing-search-engine
[23] Gillespie, Tarleton. "The politics of 'platforms'." *New media & society* 12.3 (2010): 347-364.
[24] Many arguments about the utopian and dystopian futures unleashed by social media grow out of prior contributions to media theory. Marshall McLuhan argued that core cultural technologies are able to influence the basic makeup of the human psyche. Rather than people merely using new media to send or receive communications, their attention and emotions are repolarized around it. This view positions new media as a special kind of institution that reorders the relationships of those who interact with it, whether to themselves, each other, or the outside world. According to McLuhan, the alphabet and the printing press were epochal examples of this dynamic. The former led to a psychological schism between thought and action, while the latter helped separate private life from public action (the "economic" vs. the "political") and in the process catalyzed both Protestantism and nationalism. Perhaps the most famous of McLuhan's own predictions is the emergence of a hyperconnected "global village" media network marked by tribal connections and consumption of the same mass media. It is difficult not to think of Twitter, Facebook, or TikTok in light of this prediction, with social media itself as the predominant cultural technology of the

The attention economy arose once system architectures began to support platforms on which advertiser demand outpaced the "raw" supply of user attention. This meant platform engineers could interpret attention as a scarce commodity, and apply economic theory to information management problems, specifically to maximize advertiser revenue.[25] The implications of the resultant political economy of social media platforms are now widely debated. A particularly influential lens has been articulated by Shoshanna Zuboff: *surveillance capitalism*.[26] As Zuboff portrays it, users are treated by companies primarily as sources of raw data from which insights can be gleaned to predict future behaviors and consumption patterns. While surveillance capitalism is an important dimension of the attention economy,[27] there is another, more basic one: the shift in incentives that results from unfettered attention maximization at scale. Scholars like Tim Wu have cataloged many of these shifts in terms of their effects on what is shared, who shares it, and how much there is to share: the rise of clickbait, celebrity and lifestyle branding, and disinformation, among many other effects.[28] Even without a concerted effort to monopolize user attention, and without advertising, any platform that operates at a societal scale would act as a structural catalyst for these effects.[29] Put plainly, a platform's size seems to distort how its users generate, distribute, and consume content over time.

21st century. See McLuhan, Marshall, et al. *The Gutenberg galaxy: The making of typographic man.* University of Toronto Press, 2011. See also McLuhan, Marshall, and Bruce R. Powers. *The global village: Transformations in world life and media in the 21st century.* Communication and society, 1989. Such arguments have led to provocative and compelling predictions about the cultural effects of emergent media formats. For instance, as a critic of television, Neil Postman argued that the medium's exclusive reliance on images and sound transformed all content into entertainment. Neil Postman argued that the medium's exclusive reliance on images and sound transformed all content into entertainment and education itself into "edutainment," extinguishing childhood as a distinct phase of human development. More recently, Nicholas Carr has highlighted how the internet is shortening attention spans and reducing the ability to concentrate, leading to the death of cognitive and emotional skills associated with literacy. Postman, Neil. *Amusing ourselves to death: Public discourse in the age of show business.* Penguin, 2005. Postman, Neil. "The disappearance of childhood." *Childhood Education* 61.4 (1985): 286-293. Carr, Nicholas. *The shallows: What the Internet is doing to our brains.* WW Norton & Company, 2020.

25 Lee, Ben. "Putting the 'Capitalism' in 'Surveillance Capitalism'." *Current Affairs,* 15 May 2021. https://www.currentaffairs.org/2021/05/putting-the-capitalism-in-surveillance-capitalism

26 Zuboff, Shoshana. *The age of surveillance capitalism: The fight for a human future at the new frontier of power.* New York: PublicAffairs, 2019.

27 For example, it is the case that in the search for an "epistemic monopoly" on users, social media firms may try to actively undermine or absorb rivals rather than compete with them. Facebook's decision to shut out Vine from access to its API arguably led to the latter's demise, even as it permitted the rapid growth of a certain Chinese social media service: TikTok. Meanwhile, Jillian York has highlighted the civic consequences of threats to user rights such as free speech in the context of mass data harvesting on platforms. See https://www.theverge.com/2018/12/5/18127202/mark-zuckerberg-facebook-vine-friends-api-block-parliament-documents and York, Jillian C. *Silicon values: The future of free speech under surveillance capitalism.* Verso Books, 2022.

28 See for example Wu, Tim. *The attention merchants: The epic scramble to get inside our heads.* Vintage, 2017.

29 Importantly, these effects may also arise due to multiple platforms sharing datasets or learned models, in effect generating an "algorithmic monoculture" of homogenous outcomes across distinct user populations. See Bommasani, Rishi, et al. "Picking on the Same Person: Does Algorithmic Monoculture lead to Outcome Homogenization?." *Advances in Neural Information Processing Systems* 35 (2022): 3663-3678. This dynamic can also arise sociotechnically and organizationally through a process sociologists refer to as institutional isomorphism; see Caplan, Robyn, and Danah Boyd. "Isomorphism through algorithms: Institutional dependencies in the case of Facebook." Big Data & Society 5.1 (2018): 2053951718757253.

At present, the key mechanism for platform oversight and governance is content moderation.[30] Mitigating short-term, **acute harms** alongside platform-wide rules can conflict with the financial imperative to increase and retain the number of regular users over time.[31] Yet by mediating the tension between various institutional, legal, and technical constraints, **content moderation** allows contracted employees to evaluate whether individual pieces of shared content violate platform-wide policies. It also serves as a paradigm for reconciling those policies with privacy standards, requiring ongoing evaluation of how new data collection practices interact with user expectations about the flow and sharing of information.[32] Over time, content moderation has gradually evolved into a one-stop solution for platforms to maintain credibility on how users can safely share and consume content, and for mitigating short-term, acute harms while preserving the financial imperative to increase and retain the number of regular users over time.[33]

However, content moderation is manifestly insufficient for the challenges of evaluating what can be shared at scale.[34] Even with 15,000 hired contractors (which created their own own issues[35]), Facebook's moderation team was crippled when the COVID-19 pandemic forced them to work from home or not at all.[36] Nor are firms guaranteed to support content moderation itself as a mechanism for safety. Joan Donovan's prediction, for example, that Twitter's commitment to content moderation itself was likely to weaken under Elon Musk's ownership has been confirmed, as he views these mechanisms as a tool for censorship rather than serving his customers, despite original language to the contrary.[37]

---

[30] A complete discussion of the history of (and debates around) content moderation is beyond our scope. For a deeper discussion, see the following: Grimmelmann, James. "The virtues of moderation." Yale JL & Tech. 17 (2015): 42. Gillespie, Tarleton. Custodians of the Internet: Platforms, content moderation, and the hidden decisions that shape social media. Yale University Press, 2018. Goldman, Eric. "Content moderation remedies." Mich. Tech. L. Rev. 28 (2021): 1. Douek, Evelyn. "Content moderation as systems thinking." Harv. L. Rev. 136 (2022): 526. Kate Klonick. "Of Systems Thinking and Straw Men." Harv. L. Rev. 136 (2023): 339.

[31] Lauer, David. "Facebook's ethical failures are not accidental; they are part of the business model." AI and ethics vol. 1,4 (2021): 395-403. doi:10.1007/s43681-021-00068-x

[32] User privacy has been debated and litigated for as long as online platforms have existed. For a particularly influential theory of privacy in terms of contextual integrity, see Nissenbaum, Helen. "Privacy in context." *Privacy in Context*. Stanford University Press, 2009.

[33] Hindman et al. "Facebook Has a Superuser-Supremacy Problem." *The Atlantic,* 10 Feb. 2022. https://www.theatlantic.com/technology/archive/2022/02/facebook-hate-speech-misinformation-superusers/621617/

[34] See for example Caplan, Robyn, and Tarleton Gillespie. "Tiered governance and demonetization: The shifting terms of labor and compensation in the platform economy." Social Media+ Society 6, no. 2 (2020): 2056305120936636.

[35] Newton, Casey. "The Trauma Floor, The secret lives of content moderators in America." *The Verge*, 25 Feb. 2019. https://www.theverge.com/2019/2/25/18229714/cognizant-facebook-content-moderator-interviews-trauma-working-conditions-arizona

[36] Koetsier, John. "Report: Facebook Makes 300,000 Content Moderation Mistakes Every Day." *Forbes,* 9 Jun. 2020. https://www.forbes.com/sites/johnkoetsier/2020/06/09/300000-facebook-content-moderation-mistakes-daily-report-says/?sh=44458ed754d0

[37] See: Kohli, Diti. "What Harvard's Joan Donovan thinks about the future of Twitter." *The Boston Globe,* 25 Apr. 2022.

# Towards New Infrastructure: Populations, Empirics, and Active Evaluation

Content moderation, while very important, operates at the scale of harm posed by discrete user-generated posts rather than the platform-wide interaction effects imposed by designers.[38] To be clear, we do not mean that moderation does not happen at scale – it often does, including with automated methods, such as for copyrighted material – but that limitations are imposed with respect to specific posts and not to the overall systems that platforms use to promote content. Many current and former employees believe that companies could be deploying their own internal AI tools for product optimization with respect to moderation, but choose not to in order to avoid drops in user engagement.[39] Table 2 offers an illustration of this distinction, differentiating acute and **structural harms**, which we revisit in the mechanism design section.

**Table 2:** Harm Types How They Manifest

| Harm Type | Manifestation | Examples | Regulation | Regime |
|---|---|---|---|---|
| Acute | Individual Content | Harassment / threats, self-harm, copyright infringement, pornography | Moderation | Rules Based |
| Structural | Aggregate effects of products on users in a society | Discrimination, reduced belief in institutions, mental health harms | Product Constraints | Consequential |

---

https://www.bostonglobe.com/2022/04/25/business/heres-what-one-disinformation-researcher-thinks-about-elon-musks-bid-buy-twitter/

[38] In prior work, we argued that the scale of focus and attention on content moderation – including in platforms, by regulators, and by politicians – was incommensurate with a proper framing of the challenges posed by platforms. See: Lubin, Nathaniel and Thomas Gilbert. "Social media is polluting society. Moderation alone won't fix the problem." *The MIT Technology Review,* 9 Aug. 2022. https://www.technologyreview.com/2022/08/09/1057171/social-media-polluting-society-moderation-alone-wont-fix-the-problem/

[39] As confirmed in private conversations with current and former employees.

How can we address the shortcomings of the attention economy without banning it outright?[40] An alternate regime for the continuation of privately-owned, profit-oriented platform evaluation would require a few specific components.[41]

1. **Limitations tied to population-level outcomes.** As expanded further in the Mechanism Design section, recent evidence suggests that platforms permit social comparison on a population scale,[42] leading to new types of discrimination,[43] mental health effects,[44] and declining trust in institutions.[45] These manifestations indicate that platform harms are partially (perhaps primarily) structural, rather than acute, and that the integrity of the individual user experience is distinct from – and depends significantly on – the population-level outcomes of recommender systems.[46] Other well-known social media phenomena like disinformation are likewise instances of the structural harms generated by platforms.[47] Structurally different platforms are possible. For example, Ethan Zuckerman has examined how hackathons and other new forms of participatory design made possible by platforms can open up new spaces and design experiences for vulnerable populations to be (and feel) seen. This could empower previously at-risk populations[48] but also create spaces of the "unreal," communities defined by "closed universes of mutually reinforcing facts and interpretations."[49]

---

[40] At the time of this writing, bans of social applications – especially of TikTok – have been pursued by US states and considered nationally. While such an approach may address structural harms, it is not obvious the effects of a ban are uniformly positive; for example, there may be structural benefits foreclosed by limiting platform access, even in the name of greater security or protecting at-risk populations like children. Regardless, these strategies are generally antithetical to most conceptions of market choice in all but the most extreme cases. They do not appear to be realistic (or advisable) approaches to products that have broad appeal and meaningful positive contributions, beyond whatever harmful effects they may cause (e.g. search advertising). See for example: Feiner, Lauren. "TikTok banned on government devices under spending bill passed by Congress." *CNBC,* 25 Jan. 2023. https://www.cnbc.com/2022/12/23/congress-passes-spending-bill-with-tiktok-ban-on-government-devices.html

[41] Some advocates argue that new public utilities, or antitrust, could be sufficient to address structural issues by altering platform incentives. We agree that these kinds of interventions could help place pressure on platforms to address systemic harms; however, these approaches would require difficult widespread adoption and would not be sufficient to address existing products. To take an illustrative example: new competition might induce an incumbent car company to improve the quality of its products by either addressing a poor safety record or transitioning to fully electric vehicles; yet either practice would first require evaluation (or recalls) from a trusted third party like the Department of Transportation to establish the grounds for comparison. See, for example: Zuckerman, Ethan. "The Case for Digital Public Infrastructure." *Knight First Amendment Institute,* 17. Jan. 2020. https://knightcolumbia.org/content/the-case-for-digital-public-infrastructure

[42] Guo, Elaine. "Social Media and Teenage Mental Health: Quasi-Experimental Evidence." Working paper: https://www.dropbox.com/s/5v9c6ajvmu2of8d/Guo_social_media.pdf?dl=0

[43] For example: Zang, Jinyan. "Solving the problem of racially discriminatory advertising on Facebook." *Brookings,* 19 Oct. 2021. https://www.brookings.edu/research/solving-the-problem-of-racially-discriminatory-advertising-on-facebook/

[44] Haidt, Jonathan. "Social Media is a Major Cause of the Mental Illness Epidemic in Teen Girls. Here's the Evidence." *After Babel,* 22. Feb. 2023. https://jonathanhaidt.substack.com/p/social-media-mental-illness-epidemic

[45] See Sabatini, Fabio, and Francesco Sarracino. "Online social networks and trust." *Social Indicators Research* 142 (2019): 229-260. See also Aruguete, Natalia, et al. *Trustful voters, trustworthy politicians: A survey experiment on the influence of social media in politics.* No. IDB-WP-1169. IDB Working Paper Series, 2021.

[46] Frances Haugen's Congressional testimony offers one example of how platforms harm through the feedback loops generated between users and content, rather than pieces of content themselves. See https://www.commerce.senate.gov/services/files/FC8A558E-824E-4914-BEDB-3A7B1190BD49

[47] For example: Enders, Uscinski, Seelig., Stoler (2021). The relationship between social media use and beliefs in conspiracy theories and misinformation. *Political Behavior.* And Shao, Ciampaglia, Varol, Yang, Flammini, & Menczer (2018). The spread of low-credibility content by social bots. *Nature Communications.*

2. **Iterative empiricism.** Social media platforms, in their present configuration, regularly struggle to anticipate and mitigate the social dynamics they generate.[50] The present attention economy – and the culture of unfettered optimization and growth ("move fast and break things"[51]) it has spawned – disincentivizes platforms from adopting an empirically-based regime for understanding its own effects on users.[52] Algorithmic audits comprise a potentially powerful mechanism for external accountability,[53] but are held back by limited platform access and inferior data to what is available in-house. A growing tide of evidence indicates that proactive and iterative monitoring of platform effects would be possible if companies were willing to prioritize it. The evidence available suggests that platforms are having unprecedented yet measurable outcomes on their userbases. The now-famous emotion contagion study[54] performed on Facebook raised the prospect of the platform's ability to psychologically manipulate hundreds of millions of people in close to real time. Frances Haugen's testimony confirmed that the mental health of younger Instagram users is heavily dependent on platform policies for how content is recommended.[55] Scholars like Nathaniel Persily have pointed to this evidence as grounds for providing external research access to companies' data, firstly as a basis for understanding platform effects and secondly as a basis for new regulations to prevent societal harm.[56]

3. **Real-time evaluation of product design.** Monitoring structural harms based on population-specific metrics will entail the active evaluation of platform mechanisms. This would center platform governance on regularly updated optimization constraints rather than only on passive moderation of shared content. At the heart of this alternate paradigm is the view that platforms should not strive to protect or control individual user experiences. Rather, they should focus on protecting user *populations* through an evidence-based regime of internal accountability, alongside requisite technical infrastructure to support that regime.

This qualitative shift in relevant harms, governance principles, and salient mechanisms adds up to a different paradigm for social media regulation. We refer to this new frame under the label of **accountability infrastructure** for societal-scale recommender systems. In opposition to the present attention economy, accountability infrastructure treats the effects of these systems as able to be experimentally investigated, modeled, and controlled over time. Accountability infrastructure, with its focus on the population dynamics generated by platforms, has a powerful affinity with ideas stemming from the field of public health. Within a "maximizing attention" frame, platforms are fundamental disruptors of social order, and their owners can be expected to do little more than act as referees for the most egregious types of content uploaded and shared. But with a health framework, it is possible to evaluate them as environments for **controllable interventions** on user populations, with unexpected effects interpreted as a consequence of failing to fully understand these interventions rather than as an inevitable byproduct of the cultural disruption wrought by platforms.

---

# PUBLIC HEALTH AS A WAY FORWARD

While the challenges posed by modern technology appear to be novel, there are precedents. In this section we propose **public health** as a direct antecedent, in particular the 19th century fight against epidemics.[57] Figure 1 shows how the components of public health infrastructure for disease converged over a 75+ year period.

The key lesson from this history is that societal-scale effects (and harms) require societal-scale interventions. New data sources, experimental methods, and infrastructural investments made possible the conditions for policy around public health. These historical parallels provide a lens for understanding and responding to the effects of today's societal-scale recommender systems. We deploy this lens in the following section.

**Figure 1:** Nineteenth Century Public Health: Disease Control & Prevention

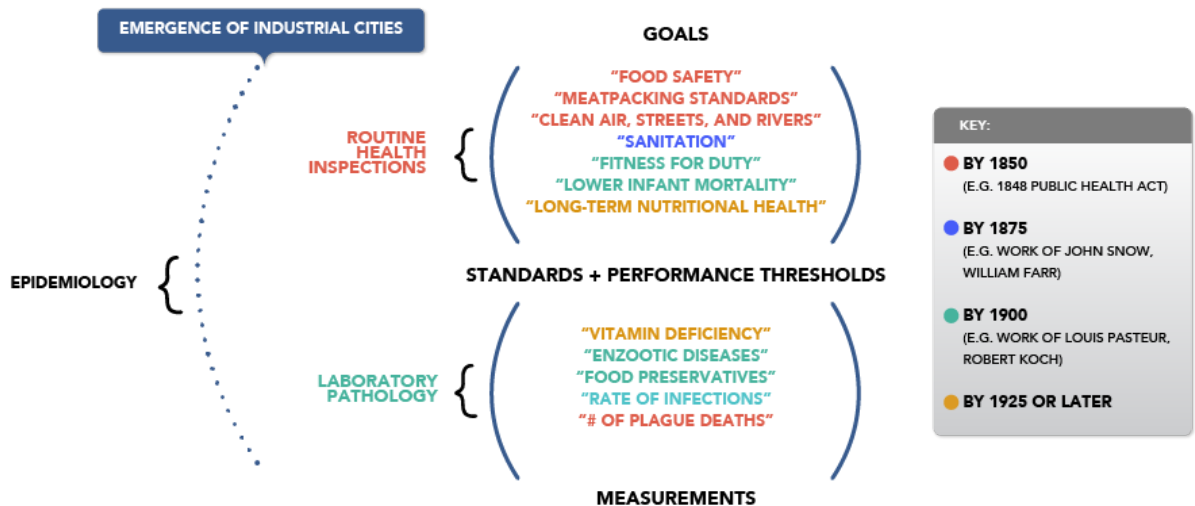

We focus on four historical transitions that directly reflect modern technology, informing analysis and the specifications of the mechanism we propose in the next Section:

1. New state and economic capacities made it possible for both unintended systemic harms to arise, and for systemic interventions on specific populations
2. New forms of evidence were mobilized that suggested causal connections unimaginable within the previous political and cultural paradigm.
3. Early policy responses comprised an attempt to make sense of these mechanisms, observing their limits and beginning to evaluate their consequences.

---

[57] Much of our content on 19th century public health comes from two sources: Rosen, George. *A history of public health*. Jhu Press, 2015. And Porter, Dorothy. *Health, civilization and the state: a history of public health from ancient to modern times*. Routledge, 2005. Other sources are cited where appropriate.

4. As evidence matured and new political coalitions arose, mechanisms were adjudicated according to standards of population health.

While modern technology developments have not mirrored this full transition – we appear to be in the second phase – we can take deliberate actions to improve our ability to forge causal connections.

## New populations and new interventions

While many of the environmental factors that led to disease outbreaks were known since antiquity, the incentives of modern states were mercantilist. Rulers tracked illnesses and tried to prevent outbreaks both because national wealth and power was tied to industry, and because sick populations made weak soldiers and an unreliable workforce. This system firmly established disease treatment as a problem of *population management* rather than *alleviating suffering*, and modern states cared about general population health even on undemocratic grounds.[58] In the transition to more modern systems, it became possible to create population-wide policies based on evidence rather than traditional forms of spiritual or political authority. Treatment of disease became an object of state investment, leading to new records of the context for disease: how many people came into poor houses, their wealth, their health, and even circumstantial evidence on how diseases spread.

This data and measurement did not add up to a theory, but it did make it possible to pose new questions about population health. And by leveraging this new data, states began to create new infrastructure and policies to curtail outbreaks once they began. However, they were ill-equipped to understand the nature of disease itself.

## Fights over causal mechanisms

Gradually, a need was recognized to understand the conditions under which disease emerged, spread, and settled within subpopulations over time. That knowledge finally came through a generation of reformers, advocates, and researchers working in the first half of the nineteenth century, climaxing with John Snow's work during the 1854 cholera outbreak in London.

Snow was the first to conclusively show that cholera infections were not a product of social station or physical location, but of unnecessary exposure to a contaminated water supply. In his investigation of the water pump on Broad Street, Snow painstakingly constructed the first natural experiment in public health, and for this work is widely considered the father of modern epidemiology.

---

[58] The emergence of population health as a topic of concern for the state lies at the heart of Foucault's influential concept of "biopower." In fact, the very term "statistics" comes from the German term for "science of statecraft" that emerged in Enlightenment Prussia. See for example Taylor, Chloë. "Biopower." *Michel Foucault*. Routledge, 2014. 41-54.

Snow's findings suggested that laissez-faire economic policies hindered rather than advanced public health, since unfettered industrialism led to squalid living conditions and societal neglect. What mattered was firm, evidence-based administrative control over health outcomes, not the autonomy of the labor market. Disease was found to be literally and figuratively downstream of economic policy, not a result of "natural" destitution. In other words, Snow showed both that the state's inaction was indirectly responsible for disease outbreaks, and that new state interventions could be leveraged to prevent or control them.

Snow's research was sufficient to accurately determine the source and means of transmission for cholera, even though he had no pathogenic knowledge of the disease itself. The biological determinants – the domain of what would later be called bacteriology or virology – were less urgent than the *social determinants* precisely because the latter reflected political choices about economic policy rather than nature. Paralleling today's efforts to combat the spread of misinformation online, modern public health policies took shape the moment that authorities could effectively intervene on how diseases were able to spread no matter the nature of the disease or the socioeconomic station of the vulnerable.

There is no better example of the transition from premodern to modern public health than the fall of miasma theory. Originally espoused by Hippocrates in the fifth century BC, miasma theory held that disease spread through bad-smelling or noxious air that traced back to fumes generated from decaying organic matter. If a neighborhood or street or river smelled bad, the argument went, it was capable of spreading disease. Major public health advocates such as Florence Nightingale and Edwin Chadwick adhered to the miasma theory – the most successful early proposals to "clean up the streets" or improve the sanitary conditions of hospitals were often justified in relation to it.[59] But in fact, miasma theory was more often a way to rationalize non-interventions based on whether specific urban areas were perceived as too "dirty" to warrant state action, rather than a means to predict how diseases spread or could be mitigated.

## Evaluating policies in light of their consequences

Though weakened by John Snow's revolutionary discoveries, the death knell of miasma theory was the rise of modern scientific tools that could isolate the mechanisms of disease. The discovery of germ theory helped show how slaughterhouses that smelled better than others could still spread disease, while putrid neighborhoods could avoid infection based on the absence of bacteria invisible to the human eye. Mortality rates rapidly declined as the older miasma theory was rejected following new technical methods for disease identification and prevention that isolated the specific bacteria and arthropods responsible for various ailments.[60]

---

[59] References to miasma theory were also very important as a basis for new arguments about the shifting relationship between public health and individual liberty, as expressed in this quote by housing reformer John Simon: "It might be an infraction of personal liberty, to interfere with a proprietor's right to make offensive smells within the limits of his own tenement, and for his own separate inhalation; but surely it is a still greater infraction of personal liberty when the proprietor, entitled as he is to but the joint use of an atmosphere, which is the common property of his neighbourhood" (28). See Stewart, Jill, ed. *Pioneers in public health: lessons from history*. Taylor & Francis, 2017.

Germ theory also cemented a shift in the political economy of health, as it interpreted persons as relational vectors of disease rather than only bearers of individual rights.[61] Modern sciences demonstrated incontrovertibly that social and economic mechanisms, rather than physical, were the basis of population health. As the specific determinants of illness became clear, the individual was reconceived, no longer as an "isolated health unit" but as a "bearer of the relations of health and illness."[62] However, for germ theory to have a concrete effect on health outcomes, its findings had to be endorsed and enforced through state policies. Germ theory was the capstone of 19th century public health because it finally answered the questions that modern states had asked for centuries about population health. It identified the causal mechanisms behind disease, justifying sanitation infrastructure scientifically rather than morally.

Modern sanitation is the institutional expression of a political commitment to prioritize the social determinants of public health.[63] Its history is the story of how a new system of control – citywide management of the water supply – made possible interventions in support of that commitment. New questions were first raised by ailments (e.g. cholera, smog) whose mechanisms were harder to pinpoint than the plagues of the Middle Ages and 18th century.

While calls for modern sanitation were motivated by the public's fear and distaste for bad smells, the point was to make public health policies *actionable*, as opposed to scientifically sound or morally desirable. A good example is the "Great Stink" in London of July and August 1858, during which the River Thames became distressingly smelly as a result of both hot weather and excessive raw sewage. In response to the Great Stink, Joseph Bazelgette was put in charge of designing a modern sewage system for a city of over two million. His innovation was to design and implement drain pipes that could redirect sewage sufficiently far away from the city that it would not flow back into the water supply.[64]

---

[60] Louis Pasteur's work in bacteriology and on the scientific (vs. state-sponsored) case for vaccinations revolutionized how policymakers thought about disease and its means of control. Bacteriology totally changed the game by allowing people to discover where disease actually came from and how to prevent their spread by accurately modeling their causes. For the most infectious and dangerous diseases, this was specific arthropods (mosquitos, fleas) that were part of an enzootic life cycle that may become an epizootic one that also contains humans (e.g. fleas→rats→humans with bubonic plague).

[61] As Porter puts it, "The real impact of infectious diseases upon the British system of public health…was not simply to simulate the creation of sanitary law, but ultimately to legitimize the power of the state to override the freedom of the individual in the reduction of the threat from diseases perceived to be preventable" (137).

[62] Porter, Dorothy. *Health, civilization and the state: a history of public health from ancient to modern times*. Routledge, 2005. p. 143.

[63] See for example Duffy, John. *The sanitarians: A history of American public health*. University of Illinois Press, 1992.

[64] Stewart, Jill, ed. *Pioneers in public health: lessons from history*. Taylor & Francis, 2017.

## New standards for health dictate acceptable interventions

Once knowledge of mechanisms and control of population interventions were both assured, the policy goals of public health became steadily more ambitious. As distinct pathologies became better understood, terminal disease prevention became less urgent. Interventions were targeted to encourage the adoption of good nutritional habits at critical moments in the life course. This shift also stemmed from growing appreciation of the social and institutional determinants of poor health. Beyond fighting specific life-threatening diseases like smallpox, reformers became interested in early childhood care, maternal support in the days immediately following birth, putting nurses in schools, and data that were *longitudinal* rather than strictly *diagnostic*. Infant mortality drastically fell, and standards of living improved even beyond the removal of terminal illnesses like smallpox or the plague. Removing disease became less urgent than ensuring healthy lives – for people to be healthy and physically fit without requiring a visit to the doctor.

Another major change was the creation of *voluntary citizen organizations* that helped combat particular diseases.[65] These were particularly effective for long-term ailments like tuberculosis that required active citizen engagement and knowledge of how to reduce disease spread through regular medical check-ups, avoiding certain poor health habits, and cooperating with medical experts who sought to implement particular lines of treatment at the regional (vs. individual) level. These interventions are opposite to those combating the plague, which would decimate a community before any of the above could take effect. In the twentieth century, *scientific nutrition* took off and served as an expert arm of preventive public health. Meanwhile the discovery of vitamins, organic molecules whose trace amounts are necessary for the proper functioning of metabolism, motivated changes to recommended diets.[66]

## Takeaways: Public health in the context of modern technology

Once matured, the four historical transitions behind modern public health result in three takeaways:

---

[65] This "voluntary health movement" was distinct from the liberal voluntarism of the early 19th and late 18th centuries, because it was targeted to particular diseases whose long-term eradication depended on a robust and informed civil society. It was not assumed to be sufficient or a general solution to public health, but a necessary tool for particular lines of "treatment" at the civic level. At stake here was the capacity of distinct demographics–generations, ethnicities, cities, genders–to decide their own relationship with particular diseases based on political and economic priorities. To be clear, this shift itself required some direct public and state investment, such as the *health education* of populations about various ailments so that their own desired relationship with it could be self-determined. That movement greatly increased the effectiveness of voluntary citizen groups for disease prevention. As Rosen writes, "What was new was the discovery of the potentialities of broad community organization as a means of controlling disease, a discovery that was to have far-reaching significance for the entire public health program" (363-4).

[66] In some cases this meant diets were *less* strictly regimented and controlled to be more appropriately diverse and balanced. For example, entire populations of East Asia, colonized by the British, were fed exclusively on polished rice and obtained crippling vitamin deficiencies. Mixed diets that included fruits and vegetables as well as proteins became more desirable than those based on crops that were easy to harvest and distribute but damaging to health over time.

1. Knowledge of the causal mechanisms for population-level harms.
2. Means to intervene in population-level environment dynamics.
3. Policy goals that define precise criteria for population health.

The 19th century struggle to prevent or mitigate epidemics benefited from a firm consensus on point #3 above: the goal was to prevent mass deaths, and the new science of statistics made it possible to evaluate census data in relation to that goal. However, point #1 was lacking – germ theory did not firmly supplant miasma theory until the start of the 20th century, following decades of debate within the scientific community followed by even more political debates about how to align policy according to the new theory in response to regional epidemics. As these elements depend on each other, the critical point is that public health infrastructure (metrics, toolkits, inspections, legislation, facilities) assumes a population-level frame for understanding harms. Their common story has an underlying tension: infrastructure must reflect a dual commitment to the scientific and normative bases of population health. The question is how to coordinate and control population interventions. The public health mindset stems from a firm rejection of strictly palliative treatment and traditional authority in favor of attention to any mechanisms that may be responsible for harms. This requires a blend of scientific curiosity, political openness, and social cohesion so that at-risk populations can be studied, organized, and protected as needed.

Crucially, the takeaways of the 19th century apply just as well to the context of 21st century platform oversight. Figure 2 reinterprets the framework of Figure 1 within the contemporary context of societal-scale recommender systems. The components of this infrastructure stem from uncontrolled feedback between users, techniques for content distribution, and high-level directives for growth. It is these types of feedback – not content itself – that generate societal-scale, structural harms.

**Figure 2:** Twenty-First Century Public Health: Recommender Systems

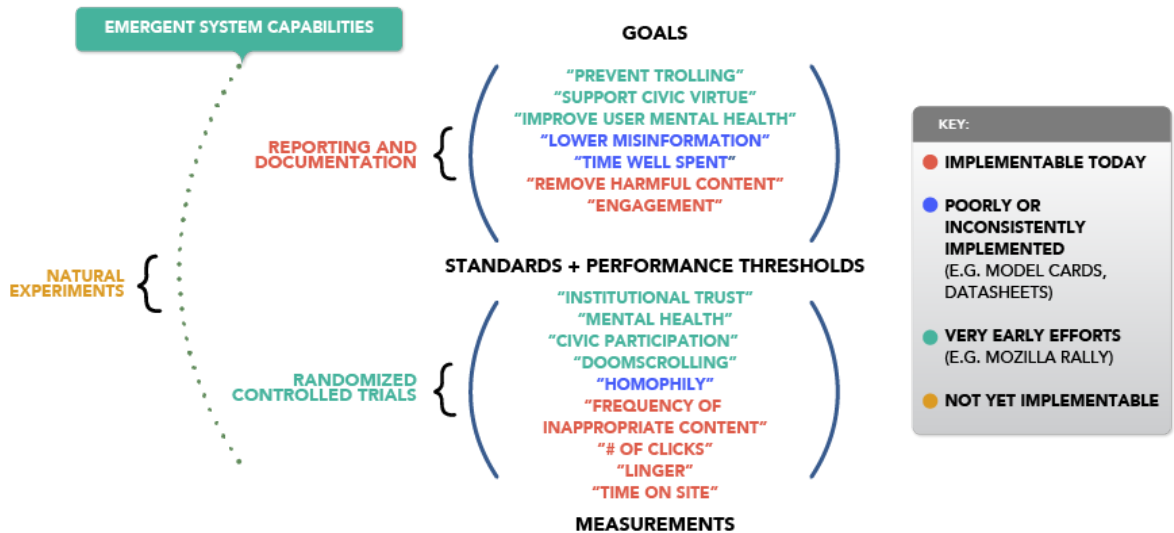

A "public health" approach to technology platforms would have to confront these feedback types according to the elements outlined above: knowledge of causal mechanisms, clear goals, and the means to intervene at scale. Posted content is akin to *water*; websites and other interfaces are analogous to *pumps*, and unintended feedback loops correspond to *risk of infection*. A public-health framework for understanding the internet would focus not on online information itself, but on how it is generated, spread, and consumed via digital platforms. These elements inform the tripartite infrastructure presented in Figure 2: regularly updated reporting, randomized control trials, and platform-wide natural experiments. Some of the components of this infrastructure are already in place; for example, platforms regularly measure according to specific metrics of concern to them, and evaluate types of content interpreted as harmful or undesirable.

However, these components are presently structured to support user engagement, attention maximization, and platform growth at all costs. As a result, the health of user populations is inevitably ignored. Integrating public health means that tech firms must both *take responsibility* for measuring the effects of their own platforms, and *be held responsible* for those effects according to public standards for safety and well-being.[67] In the present era, platform auditors are the health inspectors; ethnographers and data scientists are the epidemiologists; and system administrators are an early embodiment of the laboratory pathologist. The parallels between these roles are outlined in the table below, illustrating roles that could mitigate the externalities of the activities of platforms.

---

[67] Both points are needed for tech firms to be placed within an accountability relationship. See Cooper, A. Feder, et al. "Accountability in an algorithmic society: relationality, responsibility, and robustness in machine learning." *2022 ACM Conference on Fairness, Accountability, and Transparency*. 2022.

**Table 3:** 19th and 20th Century Accountability Infrastructure

| Accountability infrastructure roles | 19th century | 21st century |
|---|---|---|
| Reporting and documentation | Health inspectors | Platform auditors |
| Natural experiments | Epidemiologists | Ethnographers, data scientists |
| Controlled measurement | Laboratory pathologists | System administrators |

Rather than relegating these to passive checks on platform growth, public health means elevating them into critical elements of how platforms are supposed to work.

# MECHANISM DESIGN

Modern technology companies are overwhelmingly focused on growth – defined by user counts and engagement.[68] Especially for venture-backed platforms, the capacity to reach millions of users is the goal from the start, reflected in Stanford's popular "design thinking" curriculum[69,70], Peter Thiel's influential *Zero to One*[71], and Reid Hoffman's more recent *Blitzscaling*.[72] These approaches lead designers to build products that, as much as possible, generate negligible marginal costs. To implement product development in this way,[73] technology companies leverage data analytics, near-constant feedback, and reference to Objectives and Key Results (OKRs), which themselves must be specific, measurable, and verifiable via Key Performance Indicators (KPIs).[74]

Many of the largest platforms have codified this practice. Product leaders, supported by or in concert with growth marketers, are tasked with the revenue-driving part of the enterprise closest to company-wide success. Legal and policy teams – including most Trust and Safety workers – are presented as a counterweight, tasked with establishing practices and ex-post oversight sufficient for preventing abuse (inclusive of the reductions in utilization or net revenue that stem from such interventions).[75] As we have heard in many conversations with people who have operated on these teams, addressing acute harms without meaningful engagement and changes to product goals is a constant and unsustainable enterprise.[76]

---

[68] For example, Mark Zuckerberg reportedly relied on two measures for engagement – (1) how many users log into Facebook at least six out of seven days of the week, and (2) the total number of likes plus comments – to justify any significant update to the News Feed recommendation system. Hao, Karen. "How Facebook got addicted to spreading misinformation." *MIT Technology Review,* 11 Mar. 2021.
https://www.technologyreview.com/2021/03/11/1020600/facebook-responsible-ai-misinformation/
[69] "Design Thinking Bootcamp: Make Impact and Drive Growth in Your Organization." *Stanford School of Graduate Business.* https://www.gsb.stanford.edu/exec-ed/programs/design-thinking-bootcamp
[70] "Introduction to Design Thinking." *Stanford School of Engineering.*
https://online.stanford.edu/courses/xcdt110-introduction-design-thinking
[71] Thiel, Peter, and Blake Masters. *Zero to one: Notes on startups, or how to build the future.* Currency, 2014.
[72] Hoffman, Reid, and Chris Yeh. *Blitzscaling: The lightning-fast path to building massively valuable companies.* Currency, 2018.
[73] Bock, Laszlo. *Work rules!: Insights from inside Google that will transform how you live and lead.* Twelve, 2015.
[74] Doerr, John. *Measure What Matters: How Google, Bono, and the Gates Foundation Rock the World with OKRs.* Portfolio Penguin, 2018.
[75] Influenced by legal standards that prohibit nudity, violence, and abusive content, among other categories, these policies aim to protect users from discrete pieces of content that violate individual well-being. See: "How we review community content." *Meta Business Help Center.*
https://www.facebook.com/business/help/373506759931554?id=1769156093197771
[76] For further discussion of how metrics are sorely needed for the management of ethical risk, see Moss, Emanuel, and Jacob Metcalf. "Ethics owners: A new model of organizational responsibility in data-driven technology companies." (2020). Available online:
https://datasociety.net/wp-content/uploads/2020/09/Ethics-Owners_20200923-DataSociety.pdf

Despite widespread reporting and discussion of technology harms, there are few restraints on product design practices themselves.[77] There are few public assessments of where specific problems are present, how acute those problems are, or how multiple decisions create interdependencies and interaction effects. Instead of specific design specifications, we aim to scope how harm assessments might proceed to protect the public interest. These processes manifest as sequential decisions (product-based, algorithm-based, interface-based, etc.) that comprise feedback loops at scale.[78]

## What Would It Take?

To begin, we can assert a few requirements. An effective data-driven intervention must produce consistent, observable effects with defined metrics. There must be a consensus (at least provisionally) for those metrics based on specific harms, examples of which are described below. And there must be a system architecture capable of collecting and assessing the data with proper oversight but also proper data protections.[79]

Similar to the broader world of public health, a successful system of accountability requires three attributes to achieve these goals:

1. Actionable. An accountability regime must be instituted so that metrics are implementable directly within the feedback loops of product optimization cycles. Ex post assessments can be useful for auditing, but unless they are integrated into decision making, they will not enable mitigation.

2. Critical. Product designers should have flexibility and discretion over the features they choose. Because the imposition of binding limits requires a *compelling societal interest*, metrics need specific outcomes for well-defined populations of concern.[80] In practice, this very likely requires a direct connection with the health and wellbeing of a group, or the core functioning of basic societal institutions.

3. Comparable. Because the purpose of this system is to set benchmarks across use cases and for many decisions, metric outputs must be simple enough and standardized enough to enable comparisons across features/products.

---

[77] Some new efforts have begun to engage with these restrictions from user-centered design principles, such as the recent Kids Online Safety law in California: "Governor Newsom Signs First-in-Nation Bill Protecting Children's Online Data and Privacy." *Office of Governor Gavin Newsom*, 15. Sep. 2022. https://www.gov.ca.gov/2022/09/15/governor-newsom-signs-first-in-nation-bill-protecting-childrens-online-data-and-privacy/ Or see, for example: Nilay, Patel. "Can we regulate social media without breaking the First Amendment?" *The Verge,* 16. Dec. 2021.

[78] Dobbe, Roel et al. "Hard choices in artificial intelligence." *Artificial Intelligence* 300 (2021): 103555.

[79] Because these systems are iterative, initial parameters must be set prior to the establishment of the system. Ambiguity of those initial parameters is not a justification for preventing forward progress, but it is a cause for humility in initial requirements.

[80] Populations might follow traditional demographic definitions, or they might be behaviorally-determined, or defined alongside the establishment of metrics. The crucial point is that research informs where to ensure coverage of assessments, as aggregate averages may obfuscate meaningful issues on smaller groups of users.

These principles are applicable to most technology accountability regimes, whether they are internal or external to product decisions. That includes content moderation teams, which strive for their decisions to have scalability, replicability, and well-defined motivations, including for examples like CSAM[81] and terrorism material[82] where meaningful progress has been achieved.

## Approaches to Intervention

We begin with the core dynamics of a social media product, which are represented in Figure 3. We visualize this for a video system like YouTube or TikTok, although an analogous figure can be constructed for other products operating with similar feedback setups.

**Figure 3:** Core components of an online video recommender system

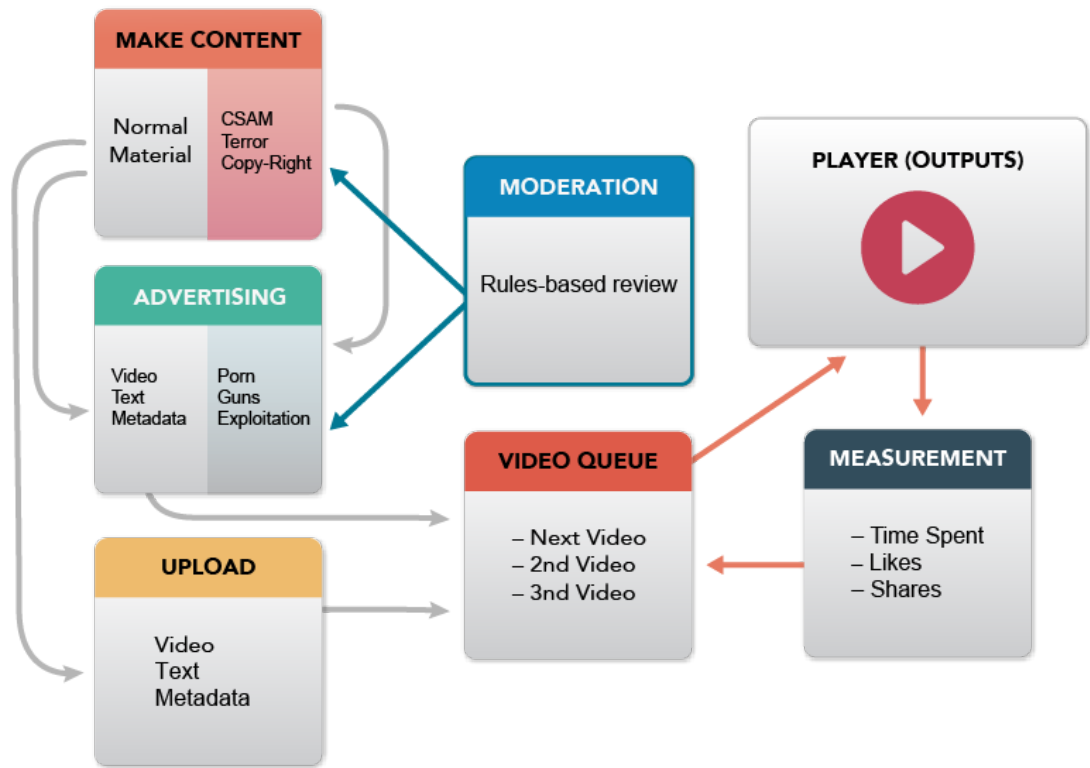

The base product experience follows a format familiar to anyone who uses social media. Content – either material shared by a user or an advertiser – is uploaded and a small subset of that

---

possible content is added to a particular user's queue. Using some type of algorithm or automated programmatic process, the "next video" in the queue is identified and the viewer is shown content. That user's response to the video is tracked in the form of an array of specific metrics which feed back into the queue, potentially updating the order of the queue so as to maximize the expected goals of the algorithm. In practice, a metric or combination of metrics associated either directly or indirectly with time spent are dominant within the current social media systems.[83]

Most major platforms have invested significantly in moderation infrastructure to intervene within this system.[84] Moderation appears upstream of the core feedback loop, and without moderation, products become saturated with abuse and harassment, advertisers become unwilling to place their brands in proximity to the ensuing feeds, and users' approval of product usage tends to be dramatically reduced.[85] Most pressingly, portrayals of CSAM and promotion of violence can lead to extreme (and often illegal) abuse.[86] Of course, where these lines are drawn has been and will remain a challenge. But events like Elon Musk's decision to fire two-thirds of Twitter's content moderators, resulting in massive increases of fraud, abuse, and incitements of violence, illustrate this challenge and its importance.[87]

Even perfect moderation is not scoped to address effects on society that arise from system effects of algorithms and the presentation of the system. Figure 4 illustrates this effect:

**Figure 4:** Sociotechnical dynamics of societal-scale recommender systems.

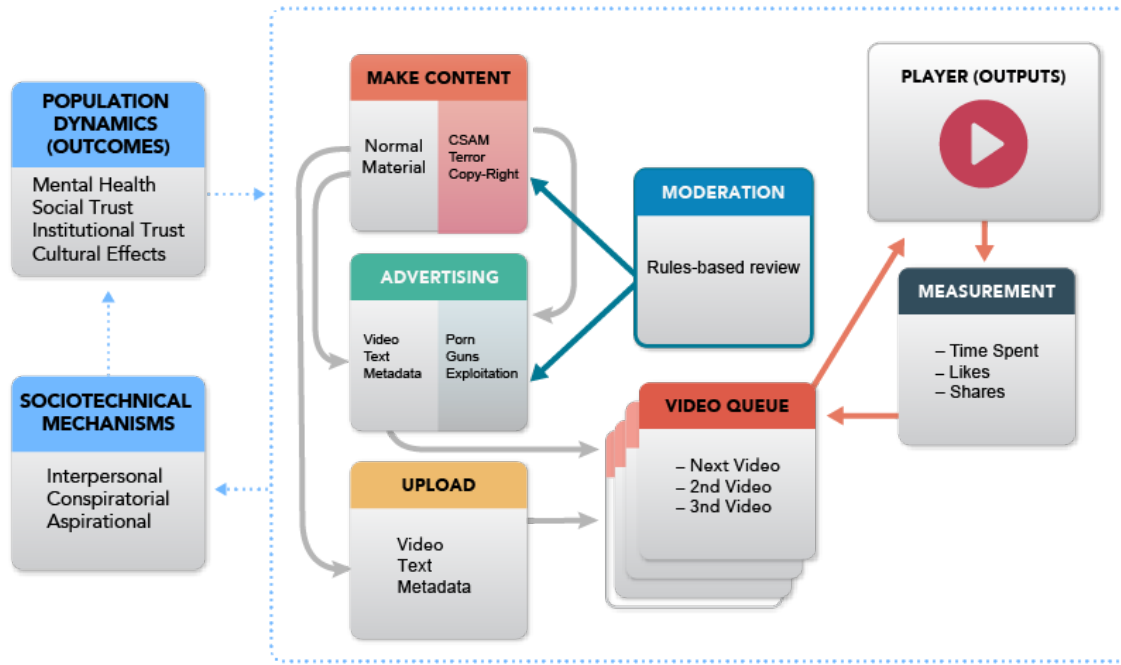

The product – inclusive of moderation decisions – can produce aggregate effects which appear outside of any individual user experience or content review process. Specifically, these aggregate effects induce various sociotechnical mechanisms which can influence individuals. These mechanisms eventually feed back onto the content queue itself, shifting it over time at scales beyond the purview of present content moderation protocols because they operate at level of feedback, outside individual pieces of content. Some of those dynamics have been hypothesized and explored by academics and researchers,[88] such as dynamics which result in: damaging interpersonal comparisons (especially by young people);[89] burrowing behavior that can result in conspiratorial bends among populations exposed to these systems;[90] and novel interaction settings for belonging and exclusion in which aspirational feelings can grow, wither, or be exploited.[91]

---

[88] Karim, Fazida et al. "Social Media Use and Its Connection to Mental Health: A Systematic Review." *Cureus* vol. 12,6 e8627. 15 Jun. 2020, doi:10.7759/cureus.8627

[89] For example: The serially mediated relationship between emerging adults' social media use and mental well-being. Rasmussen E, et al. *Computational Human Behavior*, 2020;102:206–213.

[90] For example: Foley, Jordan and Michael Wagner. "How media consumption patterns fuel conspiratorial thinking." *Brookings Tech Stream,* 26 May 2020.
https://www.brookings.edu/techstream/how-media-consumption-patterns-fuel-conspiratorial-thinking/

[91] For example: Braghieri, Luca et al. "Social Media and Mental Health." (July 28, 2022). Available at SSRN: https://ssrn.com/abstract=3919760

These sociotechnical mechanisms result in societal-level effects which a functioning society, especially a democratic one, must concern itself with. These include dynamics like the mental health of those exposed to the system, measures of various kinds of trust among nodes in the network, and the ensuing feedback loops from these effects which themselves can impact behavior on the platform. For instance, if interpersonal comparisons result in a mental health decline in a group of users, that effect can and will alter users' engagement with the platform and the platform's response.[92] If it turns out that these kinds of mental health declines *increase* usage of the platform, as in the case of social media addiction, the mechanism creates feedback for further mental health issues. Similar potential outcomes apply for other sociotechnical mechanisms.

**Figure 5:** Added constraints to monitor and control for societal-scale dynamics.

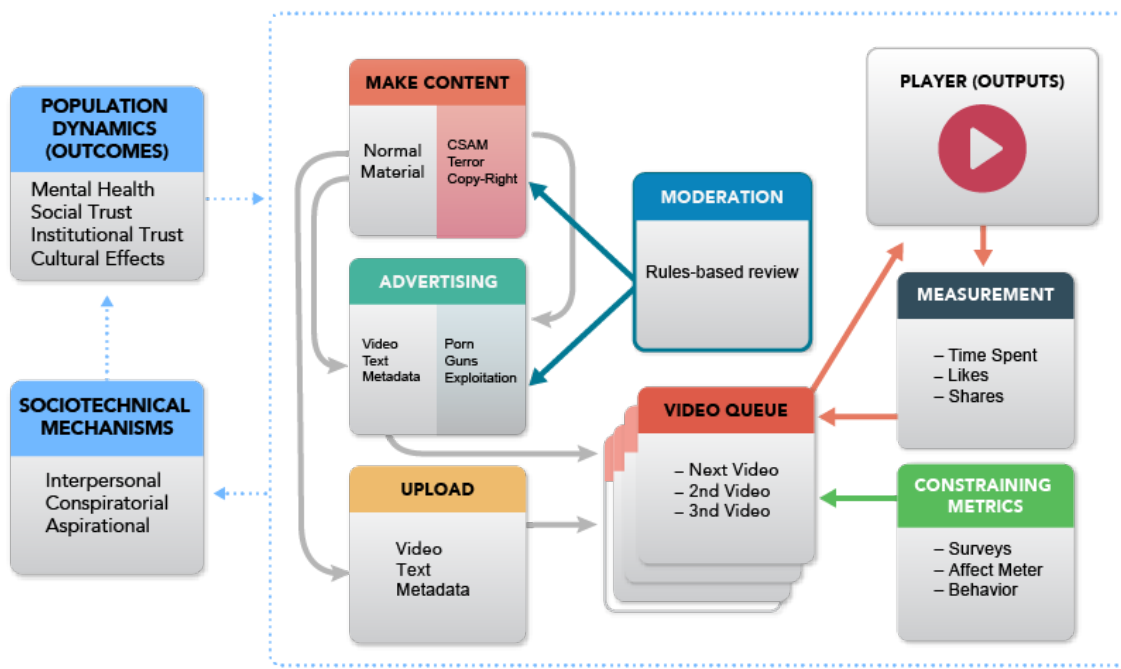

It is possible to introduce a new, additional feedback loop to intervene as a "breaker" upon these types of effects. Such a loop would appear in the system as an alternative set of metrics calculated alongside engagement which define constraints on feedback. If these constraints are properly calibrated over time and explicitly evaluated in relation to harmful outcomes, a new equilibrium could be reached. This new mechanism would make it possible to actively balance the interests of the public with the private interests of the companies' metrics in terms of the behavior of the platform.

---

[92] Wetsman, Nicole. "Facebook's whistleblower report confirms what researchers have known for years." *The Verge,* 6 Oct. 2021.  https://www.theverge.com/2021/10/6/22712927/facebook-instagram-teen-mental-health-research

With this formulation of the problem, we can also portray this system challenge as a three-part feedback cycle, illustrated in Figure 6:

**Figure 6:** Accountability infrastructure for societal-scale recommender systems.

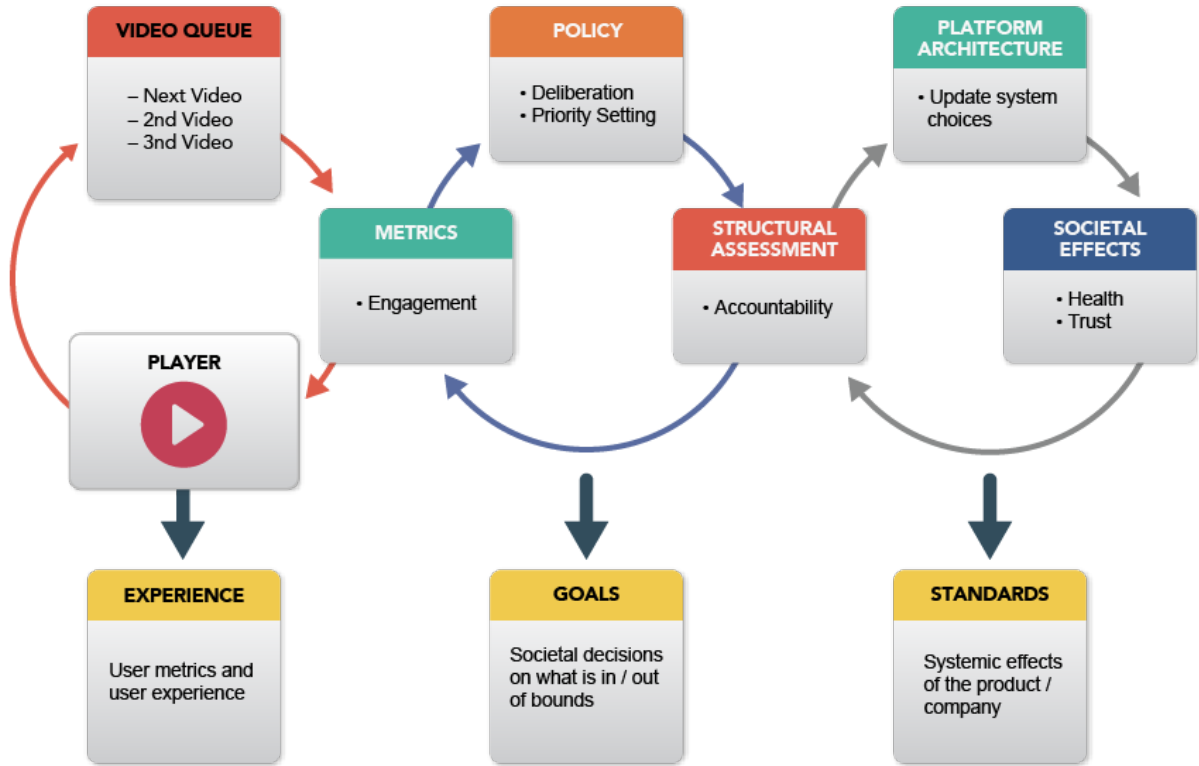

We believe these three feedback cycles define an effective accountability infrastructure. At present, major platforms maintain the "experience" loop and a (largely implicit) "standards" loop. Platforms actively assess the experience of users and are concerned a great deal with how users' perception of products impact their decisions to use the product.[93] Ultimately, this appears in engagement metrics and assessments.[94] The "Standards" loop also operates, even if it is poorly understood due to lack of direct systems of assessment. We see circumstantial evidence for

---

harms on at risk populations, such as teenagers[95] or inducements toward radicalization.[96] Without defining the middle "goals" loop – one by which some mechanism bounds on societal effects are set directly – both causal measurements and alternative deliberation for product design will remain extremely difficult.

External regulation is not needed to implement mechanistic interventions. Internal product leaders and ethicists in companies like Google and Facebook can and have made meaningful interventions without public scrutiny, often over the interests of product growth.[97] But we know from many sources that this type of intervention is often fraught for those who undertake it.[98] Many of the individuals who have worked on these projects subsequently left their companies (not always voluntarily[99]) while calling out the obstacles posed by internal as opposed to external regulation.[100]

The public health model suggests the need for an administrative agency setting targets and conducting inspections. We remain hopeful that product leaders in industry will adopt a framework like the one described here as a means towards that end. Even without an immediate product implementation, planning for a future evaluation and incorporating a commensurate set of evaluation systems would constitute a major design shift. Still, we believe the case for oversight is incredibly strong: actual requirements for changed behavior would be localized only to cases where private actors' decisions are having strong, observable, preventable, negative effects on other cohorts of society, and those harmful externalities can demonstrably be reined in.[101]

Such a framing begs the question, then, if it is possible to assess systemic effects at all, what legitimate metrics should we pursue under this regime? In evaluating specific implementations, we can return to the initial setup questions posed at the start of this section and ask now what it would take to build a system capable of assessing and adjudicating upon metrics that are

*actionable*, *critical*, and *comparable* while also operating at the scale of system architecture as a whole.

It will fall on tech companies to enact these metrics on their platforms and factor them into optimization criteria explicitly. But to be clear, this does not mean that companies can exercise discretion over the metrics themselves. Our point is the opposite: the criteria for what metrics count as actionable, critical, and comparable must be drawn from legitimate sources that clearly reflect the will of affected publics. The struggle to coordinate these projects – the political articulation of public problems, and their economic measurement by private companies – is a necessary tension and distinctive feature of public health.

## Population Dynamics and Sociotechnical Mechanisms

Establishing an implementable system requires participation from a range of stakeholders and experts.[102] For now, we focus on the areas where there is the clearest consensus on the need for intervention, an appropriate approach for two reasons. First, there is an obvious bias towards political inaction. But second, systems designed to address one harmful population dynamic may also help address other harmful dynamics as well because different negative outcomes may share the same underlying sociotechnical mechanisms. For example, a system designed to mitigate trust reduction stemming from a conspiratorial mechanism might have a similar mitigating effect on mental health stemming from the same comparative mechanism.

Metrics operate at the level of the relevant group, rather than the individual, recognizing that overall averages can mask meaningful differences in subgroups where the worst structural harms will often be localized. This is highly analogous to public health: for most public health questions and interventions, the "community risk" of analysis is neither the individual nor all members of society, but rather a considered risk pool.[103] So long as the pool is well defined, that definition may be based on physical proximity, demographics, behavior, or most any other behavior or attribute.

### Mental Health & Social Trust

Two categories of harmful public dynamics are particularly strong initial candidates for this regime: **mental health effects**, especially for at-risk populations like teenagers, and **measures of interpersonal trust**.

---

[102] While technologists and regulators undoubtedly are central to this process, an effective regime might include participation from diverse fields such as public health, mental health, child development, political science, philosophy, sociology, and law enforcement.
[103] See: "Human Health Risk Assessment." *Environmental Protection Agency*. 26 Jul. 2022. https://www.epa.gov/risk/human-health-risk-assessment

There are strong preexisting beliefs in the importance of both mental health[104] and trust,[105] and strong existing support for action to advance each of them.[106] Mental health has been the most consistent and dominant thread of oversight in the US Congress[107] and has been the subject of actual legislation.[108] It has been adjudicated in courts,[109] drawing direct political action, and also is the subject that public research suggests the American people (across parties and ideologies) care the most about related to social media.[110] Trust, by contrast, is less frequently associated with technology systems, but reductions in standard measures of trust have been decried by academics and groups of all stripes, including with reductions associated with rises in violence,[111] reduced economic output,[112] and reductions in social capital generally that are highly connected to well being.[113]

These categories have clear appeal: both are apolitical and for both no ideology or political group can credibly represent them as anything other than in service of the public interest. Second, they are particularly attractive when compared to content review via moderation, which immediately triggers regulatory questions about restricting access to or banning content (see Discussion). Lastly, despite limits in direct access to private social media, prior research has already convincingly demonstrated that social media platforms can and do have systemic effects on these measures.[114]

What do we mean by terms like "mental health"? Much research has examined these attributes in a range of contexts and considerations. In our review of existing literature and conversations with experts, we have seen a number of possible approaches[115] – and yet there does appear to be sufficient consistency and similarity that research-grounded provisional definitions can be established.

Our preferred method for metric selection would be a procedural one, leveraging a body of experts who would be responsible for evaluating end results. Nonetheless, we believe it is important to outline provisional options for the two objectives proposed here, as there are a number of suitable benchmarks.

For mental health, most measures reduce to assessments of "low mood, lack of interest, appetite, and sleep"[116] with well-defined batteries of metrics which define low mood[117]. One approach that appears to have consistent use is survey benchmarks of between two and 20 questions. These include the "Patient Health Questionnaire -2"[118] which includes just two questions or its eight-question cousin.[119] Similar surveys exist for anxiety (also ranging from two questions to longer form)[120] and stress.[121] Other more generalized standards include the mental health

Continuum Short-Form (MHC-SF) assessment.[122] According to this test: "the short form consists of 3 emotional well-being items (reflects hedonic well-being), 6 psychological well-being items, and 5 social well-being items (when combined, reflects eudaimonic well-being). Contingent on response quality (and honesty), these response options assess the frequency with which respondents experience each symptom of positive mental health. This scale also provides a flourishing and languishing mental health indicator based on these three subscales."[123] Questions include assessments of happiness, assessments of positive functioning in society, with established benchmarks for "flourishing," "languishing" and "moderate" mental health.

More specific metrics have also surfaced in conversations with academics and public health officials focused on addressing eating disorders. There is a quite widely accepted standard diagnostic test called SCOFF, a five-item measure to gauge levels of risk.[124] While professionals in this field have shown the limits of these types of diagnostics,[125] the best current research suggests that there are high correlations between social media usage and direct measures of depression. The same benchmarks used by those researchers could be deployed natively within a platform, including breaking down results by developmental cohort akin the age-segmentation demonstrated in recent scholarship (which shows the biggest effects among 11-13 year olds identifying as female, 14-15 year olds identifying as male, and 19 year olds of all genders).[126]

For trust, the so-called "general trust question" asks people whether they believe "most people can be trusted" or whether "you can't be too careful when dealing with people."[127] Other variants exist, but groups such as Pew Research[128], as well as a range of academics[129], have found stability and consistency with these formulations over time. We believe initial diagnostics can be established by experts in these fields to enable at least provisional research, taking advantage of prior methods and standards for evaluation that have been validated for other purposes.

More specifically, Pew has asked a short battery of questions for years, which are quite similar to the academic literature. These include:[130]

- Generally speaking, would say that [most people can be trusted / most people can't be trusted]
- Do you think most people [would try to take advantage of you if they got a chance / would try to be fair no matter what]
- Would you say that most of the time people [try to help others / just look out for themselves]

These formulations are all survey-based. Below, we discuss alternatives to surveys, although we believe some formulation of these kinds of established norms will be necessary, at the very least as a training set for other correlated metrics.

## Feedback System Measurements: Possible Mechanisms for Randomized Controlled Implementation

Once goals for assessment – i.e. mental health and social trust – have been defined, assessments can be added to randomized controlled trials.

There are three questions of this potential system:
1. How, and in what situations, can suitable data be sourced from the platforms as part of their existing (or potential) product evaluation processes?
2. How, and by who, will tests be conducted, taking into consideration user privacy and, potentially, an institutional review board (IRB) or an equivalent's review?
3. What is the mechanism to update or add metrics or create new areas for inquiry/evaluation, consistent with our fallibilist approach?

We consider each of these questions in turn.

### Technical Systems

As mentioned, both mental health and social trust research typically takes the form of survey research. While poorly-designed surveys have limited diagnostic value, they nonetheless remain the gold standard for many difference-in-difference methods, including randomized control trials and longitudinal analysis.[131,132]

---

[130] Rainee, Lee et al. "Trust and Distrust in America." *Pew Research Center,* July 2019. https://www.pewresearch.org/politics/2019/07/22/the-state-of-personal-trust/

Essentially every scaled digital platform already surveys users or has the capacity to do so. Products like Facebook and YouTube use surveys in their standard evaluation of product development,[133] user studies,[134] and to support advertising.[135] Public reporting has described how these companies use surveys associated with randomized controlled trials of new product features to evaluate how and whether to scale deployment. For the most part, these uses are highly focused on perceptions of the product associated with use and engagement, or with advertising. But in at least some cases, reporting has indicated that the results of surveys have been used in decision making to scale back harmful effects of products.[136] Of course, because these decisions operate outside the public eye and without external evaluation, we must rely on investigative journalism and the partial views exposed in leaked documents to understand how and what happened.[137]

Product groups at major companies already utilize holdout experiments to gauge the engagement effects of their product decisions. As a result, whether or not surveys are utilized for evaluation, there are well-defined test populations generally available for evaluation. Some of these results have been made public via leaks and disclosures, although a full accounting is not public. Discrete experiments on products like newsfeed have illustrated, though, that tradeoffs of the type described here are already made in at least some cases for issues like misinformation and hate speech.[138]

Because these types of surveys are already used as a matter of standard practice, alternative metrics could often be built directly into existing systems.[139] For other tools which do not use these systems at all (at least publicly, we are unaware of them for products like TikTok, for example), we believe that implementing an architecture for a set of simple diagnostics would still be relatively straightforward. Because many of these updates and changes are small, in most cases these evaluations would operate at the level of aggregate changes by product groups, or even a company as whole, over longer periods (e.g. quarterly trackers to match the internal KPI cycles of many companies).[140]

At the scale of products like Facebook, YouTube, or TikTok – with millions of daily users – there are also potential opportunities for more sophisticated statistical modeling of behavior to understand effects on subgroups over time. Tools like multilevel regression with poststratification could be used to evaluate estimates for pre-defined populations that might not source enough data on their own to garner meaningful measurements.[141]

Our point in making the case for surveys as a key component is not to argue for their methodological primacy, nor to advocate for their sole use. Rather, we believe they offer an easy and legitimate onramp for most any large product with a UI placed in front of defined user cohorts. There are other methods that might prove even better diagnostically. For instance, tools like the Photographic Affect Meter (PAM) could likely be implemented quite directly in the UI of a social media product to evaluate mental health effects alongside other methods.[142] PAM is an open source system that has been used in clinical trials, behavioral research, and workplace assessments, offering a simple mechanism for assessing user affect via the affirmative selection of one of 25 images which (in aggregate and after validation) offer diagnostic value. Affect is likely a critical evaluation measure for mental health – and with the resources of large platforms, additional research and tooling could be developed to expand the approach to pick up other

---

See for example Borchers, Andrea T., et al. "The history and contemporary challenges of the US Food and Drug Administration." Clinical therapeutics 29.1 (2007): 1-16.

[140] A harder question relates to defining adequate power calculations for at-risk user segments. At the level of abstraction described here, we cannot parameterize every population upon which harms will operate – and we expect that many relevant groups will not be measurable. For instance, we do not believe this regime will be suitable for directly addressing groups which associate with the rise of violent extremism – many of whom, of course, would be unlikely to participate in research. At the same time, we do not see this limitation as an issue with the core motivation for the approach: these types of cases in their extreme forms are precisely where moderation needs to step in, because harms will manifest in ways that are acute and where an intervention is facially required. Even still, when groups like DARPA and the broader military are asking what efforts can be taken upstream of radicalization to make such behavior more unlikely,# efforts to foster social trust – or, more specifically, to prevent the erosion of trust due to exposure to a product feature – are important potential responses. See: https://idstch.com/geopolitics/darpa-incas-will-exploit-social-media-messaging-blog-data-to-track-geopolitical-influence-campaigns/

[141] Ghitza, Yair. and Andrew Gelman. "Deep Interactions with MRP: Election Turnout and Voting Patterns Among Small Electoral Subgroups." American Journal of Political Science, 57: 762-776. 2013. https://doi.org/10.1111/ajps.12004

[142] Pollak, John et al. "PAM: a photographic affect meter for frequent, in situ measurement of affect." CHI '11: Proceedings of the SIGCHI Conference on Human Factors in Computing Systems. 725-734. May 2011. https://dl.acm.org/doi/10.1145/1978942.1979047

signals. Specifically, we believe visual cues could likely be developed and validated for interpersonal dynamics as well, thereby offering an on-ramp for measures like trust.

Extensive research has connected behavioral information associated with smartphones and fitness devices to health. This data (like sleep patterns and heart rates), generally does not fall under HIPAA standards of scrutiny, and would likely be able to source large and representative populations of opt-in users. Similar to surveys or tools like PAM, populations of users exposed to specific products would be evaluated by comparing (large/representative populations) physical differences on these metrics. For subgroups like kids, reductions in sleep may prove to be causally tied to reductions in subjective wellbeing all by itself.

Perhaps most concretely, on-platform metrics have been associated with measures of health and interpersonal dynamics associated with measures like trust.[143] The social contagion experiment at Facebook illustrated quite concretely risks associated with these methods in coordination with academic research[144]. To the extent such effects can be connected directly with structural harms suitable for research, similar methods would be legitimate candidates for assessment as part of this regime. We expect that they would, but especially if used alongside externally-validated methods along the lines of surveys or some other non-engagement proxy.

A robust regime would likely combine these techniques, and others surfaced from internal teams at companies or from scholarly experts. In so doing, a successful methodology will strive to balance simplicity with community values,[145] reliability for specific subpopulations over time,[146] sensitivity to the level and distribution of health in the population,[147] and incorporate large sample sizes for evidence of population health improvement.[148] Each of these procedures, though politically impartial, would help mitigate threats like misinformation and distrust of institutions, since those effects are highly correlated with divisiveness and anti-social conflict.[149]

---

## Which Products Are Included?

Technically, the simplest strategy is to employ standards for evaluation alongside existing A/B testing protocols. Because essentially every product on scaled platforms is already deployed to small cohorts of users to evaluate behavioral response (generally to boost engagement, time spend on site, or some other monetizable dynamic), a policy could require that diagnostics (e.g. surveys, PAM, etc.) be included in assessments of A/B rollouts of scaled product changes. This would enable a direct measure of the effect of the new feature. Changes might require large datasets to evaluate, and might be subjected to only a subset of tests which could plausibly result in material effects,[150] with the goal of such a regime being to shift the default procedures. If sufficient samples are impossible in the course of an A/B test, a follow-on evaluation could be deployed for live testing compared to controls after scaled deployment, similar to sequential pharmaceutical trials.

One benefit of associating a regulatory structure like this with actual product development is that aggregations of engineering changes may be combined. For instance, if non-trivial changes to a core recommendation algorithm were implemented quarterly, a comparison test might incorporate all changes made during that quarter and compare to the previous version rather than an evaluation of any single product feature modification. Such a shift would not require any individual change to "pass" the test; rather, it would ensure that the aggregate effect does not move in a negative direction over time. This strategy would also allow companies to make tradeoffs in their design processes, rather than over-regulating any specific decision.

In practice, and in consideration of conversations with existing product teams in companies, we believe there are three product scenarios for practically implementing these kinds of methods to gauge "meaningful" product updates. Each of these approaches would be suitable in different contexts, or could be used in conjunction, including:

1. <u>Direct, large-scale rollouts of new products/features:</u> The direct evaluation of a new rollout would be the simplest direct implementation of this protocol. In cases where a new production is being deployed – e.g. a completely new object within a feed, or a brand new instantiation of a recommender system – extensive testing of the product is certain to occur. For products like Facebook, these tend to take the form of small-scale rollouts in discrete geographies to a small share of users, followed by escalating deployment to larger and larger user segments based on success (on an engagement/profit basis).

---

[150] A scaled, or "meaningful," product change under this regime is one that can cause structural harm. By definition, that means a measurable effect to a population and not only an individual within the defined metrics (e.g. mental health). This requirement necessitates a focus on architecture questions which could plausibly reach significance in an assessment, differentiating between changes like ranking algorithms (which should be evaluated) and changes in button color (which should not).

2. <u>Continuous integration of features operating at the level of the product team:</u> More often, product and design teams iterate upon existing products, pushing incremental developments that are continuously integrated into a live environment in batched groups. This kind of approach may be more efficient for engineering and management reasons, and no reasonable enforcement regime could disrupt these processes; similarly, many of these updates are not of the scale that individual modifications are likely to manifest in statistically significant outcomes in an accountability regime. Nonetheless, in aggregate, these changes are important – so important that teams almost always already include holdout audiences to evaluate the aggregate effect of their work. Under this regime, impact metrics would be evaluated alongside engagement metrics among both holdout and treatment audiences, gauging relative effects of the full block of rollouts at the team level. These could be deployed at the normal cadence of evaluation of the team if faster than quarterly, or quarterly otherwise.

3. <u>Population holdouts over the full product offering with periodic assessments:</u> This is the most universal, and most easily implementable recommendation. It is also a clear starting point for any product that does not have built-in hold out definitions within product development processes, including most small product deployments. In this case, the product would be judged in its totality, not attempting to differentiate between feature changes; the upside of this approach is that it is universal – but the downside is that it would limit insight into where/how particular product choices impact outcomes. In the event a negative result were to be found in this regime, the response would be a product rollback to the last tested version of the product. Audience definitions would be organized identically to the team-level holdouts – except that rather than matching a particular sub-audience of users to that product, the holdout audience would be selected from the full user base.

In each of these cases, special consideration would need to be taken for subgroup analysis of at-risk populations, and particular steps would need to be taken to ensure data coverage of risk factors associated with the product. Specifying these details is impossible outside the context of a specific product. However, we can be confident that audience coverage for at-risk populations would likely demand oversamples for inclusion. These may include communities of color, queer communities, or religious minorities. Geographic oversamples would also likely be necessary.

## For which companies?

Any media product capable of causing structural harms on a scaled population is potentially suitable to this structure, which definitionally requires a scaled userbase. Regulation in the EU[151]

---

[151] "Questions and Answers: Digital Services Act." *European Commission*. 25 April 2023. https://ec.europa.eu/commission/presscorner/detail/en/QANDA_20_2348

and proposed regulation in the US[152] has set varying numbers of users for triggers of regulation. Our preference would be for an escalating set of requirements for larger userbases, although simple rules like the 10% of population which constitute large platforms in the EU could be suitable for a simpler structure. One option for a regime like this might look like the following:

**Table 4:** Requirements by Product Usage Scale

| Number of Monthly Users | Requirement |
| --- | --- |
| 1 million | Submitted plan for metrics and methods for evaluation of potential structural harms |
| 10 million | Consistent data collection on potential structural harms |
| 50 million | Quarterly, enforceable assessments on product aggregate effects on structural harms, with breakouts for key subgroups |
| 100 million | Monthly, enforceable assessments on product aggregate effects as well as targeted assessments of specific product rollouts for any subproduct used by at least 50 million users, with breakouts for key subgroups |

Such a policy would avoid overly-burdensome product requirements for startups that might otherwise stifle innovation and competition (i.e. products with less than 10 million users), but give companies context for when and how they would need to account for growth.

A "product" does apply to systems and architecture of media outside of just "traditional" social media like YouTube/Facebook/Instagram/TikTok. While those services, or any future additions, would be the likeliest to have direct questions with their products due to their attention-oriented business models, other services would also be potentially subject to these kinds of assessments. For instance, the recommendation algorithm for a service like Netflix – to be clear, *not the content included on the service*, akin to a product like TikTok – would also be subject to assessments of potential structural harms, as would the integration of generative AI into massive search engines and interfaces. Recommendations made by news or news-adjacent organizations might also be subject to this review (addressed further in the Discussion) – enabling a way to have a business assessment of the effects of these kinds of services without in any way limiting the decisions and publication choices of any newsroom or media service.

---

[152] Klobuchar, Amy. "S.2992 - American Innovation and Choice Online Act." 2 Mar. 2022. https://www.congress.gov/bill/117th-congress/senate-bill/2992/text or Coons, Christopher. "S.5339 - Platform Accountability and Transparency Act." 21 Dec. 222.  https://www.congress.gov/bill/117th-congress/senate-bill/5339

Critically, because the standard for causing a structural harm of the kind outlined in this section is so high,[153] we expect most products to quite easily pass an assessment.

## Reviewers, Administration, & Reporting

As mentioned at the start of this Section, the system described here does not specify where or how adjudication occurs. We believe that these approaches to accountability systems could be implemented by companies themselves without external requirements; we also believe these methods are suitable for the sorts of government regulation currently imposed for product safety for physical goods.

Whether in-house or managed by a third party, an effective and legitimate regime will have certain implementation and oversight needs. Steps to build such a regime would include:

---

**Table 5:** System Requirements for Administration & Reporting

|  | Requirements | Descriptions |
|---|---|---|
| Step 1 | Goal Setting | Prioritize structural harms according to public urgency and/or readiness of standards for evaluation. Mental health (especially for teens) very likely has already reached this status. |
| Step 2 | Metric Definition | Define an index of metrics, including standards for what effect sizes are large enough to trigger intervention, and standards for which subsamples must be included in an evaluation. |
| Step 3 | Usage Rules | Set a floor for product usage to be subject to review, potentially with an escalating set of parameters tied to scale. |
| Step 4 | Measurement tools | Implement product(s) capable of measuring the index, such as surveys, PAM, external health devices, or in-system behavior triggers. |
| Step 5 | Team Structure | Identify a body of evaluators independent from product developers to perform analysis, with an oversight system. |
| Step 6 | Reporting Rules | Define transparency and reporting requirements that ensure compliance in implementation. |
| Step 7 | Mitigation Procedures | In the event mitigation steps are needed, create a tracking system to ensure sufficient intervention. |
| Step 8 | Escalation Procedures | Provide an escalation system in the event that mitigation steps are not followed or are not sufficient to address the structural harm. |
| Step 9 | System Revisions | Provide for a system to update the regime inclusive of triggers for intervention, product solutions which define the index, or the addition or removal of top-level structural harms for inclusion. |

For native implementation within a company, the key step is independent evaluators. To avoid conflicts of interest, a regime requires an external assessment team without a monetary incentive to evaluate data, likely from academic or non-profit entities with missions tied to the public interest. The proposal put forward by Nate Persily for a system of external access and review of data would be directly suitable for this purpose.[154] The practice organized here differs in the scope and methods of problem definition and timing of data collection, but not in systems of data protections. Specifically, under that proposal, only approved research questions (such as the ones outlined here) would be implementable, and only by approved and credentialed researchers.

---

[154] Wright, Tara. "The Platform Transparency and Accountability Act: New Legislation Addresses Platform Data Secrecy." *Stanford Cyber Policy Center,* 9 Dec. 2021. https://law.stanford.edu/press/the-platform-transparency-and-accountability-act-new-legislation-addresses-platform-data-secrecy/ See also: Coons, Christopher. "S.5339 - Platform Accountability and Transparency Act." 21 Dec. 222. https://www.congress.gov/bill/117th-congress/senate-bill/5339

The implementation of this type of system in conjunction with our problem setup definition is functionally very similar to pharmaceutical trials suitable for public health usage. Users and direct accessors of data are both blind to the research question and product correlation; only a small research group has after-the-fact access to data benchmarked by a control group; and there is oversight by an accredited body to ensure compliance and privacy protections.

## Privacy Systems & Protections

As discussed in the prior section, the privacy protections envisioned here would be modeled on a combination of prior proposals and standard practices for randomized controlled trials used in health regulation. While it is possible that the adoption of our framework would lead to a revaluation of privacy in the context of platforms – and such revaluation may be desirable – existing standards are an adequate starting point for accountability infrastructure.

Several protections for data access can be built into implementation, relying on the focus on aggregate experimental results rather than data inputs:

- Raw data does not leave controlled access points native to the technical infrastructure of the companies
- The only access to raw data would be those people (including third party reviewers) who are directly working on a specific evaluation
- Blind participation in data collection by those users in relevant groups, other than prior notification that they are contributing to evaluations only usable to improve the system
- Analysis assessments tied to population exposures (including relevant subgroups) to a given product, with no personally-identifiable information associated. For transparency reasons, these assessments would be published within an academic/health review, similar to a health trial, ultimately with a public summary of the results.
- In the event any sample data or other detailed descriptions of subgroups is needed to be made public as part of such a procedure, implementation of differential privacy protections by default in all cases.

These provisions are not meant to be a comprehensive characterization of privacy challenges posed in the analysis of health data. Rather, the intention here is to rely on existing practices as they are used elsewhere, relying on credentialed third party review as the key differentiator – with no PII transferred outside of the analysis itself. Data practices here are parallel to those under proposals like the Platform Accountability and Transparency Act,[155] but with more direct and specified summary deliverables for public metrics.

---

[155] Ibid.

# DISCUSSION

Our intervention methodology, with mechanism design at its core, requires a clear statement of what the system is intended to do. Its specification must include an articulated goal (or, in the parlance of mechanism design, an objective) that conveys the intended outcome of the system as a whole. Despite its many measurement problems, 19th century public health had outcomes whose stakes and goals were clear (minimize deaths, prevent epidemics). This made it comparatively easy to evaluate interventions even if precise knowledge of mechanisms was not available.

The promises and limitations of mechanism design raise many unresolved questions about how to govern platforms. An obvious current example is TikTok, famous for its "For You" algorithm that automatically recommends content to the user without them having to directly engage or click on anything. What makes the "For You" algorithm so magnetizing (and addictive)? Every indication is that the algorithm in question is not more sophisticated or "intelligent" than those that power the recommender systems of the traditional Facebook app or Twitter. The likely explanation is the tight coupling between TikTok's algorithm and its user interface – which is incredibly well-designed.[156] The 15-second minimum and 60-second maximum on uploaded video content perfectly matches the length of short-term memory (around 15-30 seconds), effectively turning the user into a passive content cycler.[157] Influencers and media personalities have flocked to the platform to take advantage of this, making it a premiere outlet for user-generated content. Other major platforms have emulated TikTok's format (YouTube Reels, Instagram/Facebook Stories), capitalizing on large userbases exposure to new short-form products, but within a less seamless interface. TikTok combines what made Vine so captivating and Facebook so encompassing: the ability to "channel surf" across one's own social network atop an effervescent foam of shared content. While TikTok's recommendation algorithm does permit this affordance, it is the alchemical combination of algorithm, interface, and social conditions that generates unfettered engagement (and potential structural harms).

Because of this complexity, proactive mechanism design will not automatically address the cognitive, cultural, or sociological dimensions of societal-scale platforms. But it will make those dimensions more tractable from a policy standpoint, raising important questions that cannot be fully addressed within the current, growth-oriented paradigm. In this section articulate some key questions, sketch their outstanding features, and clarify how or whether the principles of mechanism design could help address them.

---

[156] Matsakis, Louise. "Tiktok Finally Explains For You Algorithm Works." *Wired*, 18 Jun. 2020. https://www.wired.com/story/tiktok-finally-explains-for-you-algorithm-works/
[157] Cascella M, Al Khalili Y. Short-term Memory Impairment. [Updated 2022 Jul 21]. In: StatPearls [Internet]. Treasure Island (FL): StatPearls Publishing; 2023 Jan-. Available from: https://www.ncbi.nlm.nih.gov/books/NBK545136/ Note that these effects are magnified by TikTok's tendency to aggressively push videos in front of random users to see which ones attract interest. Every video is shown to at least a few other users (something TikTok can do because people interact with an endless feed rather than by searching). As a result, lots of videos have short virality, rather than a few videos dominating for a long time. We are indebted to James Grimmelman for this point.

## Towards an Objectives-First Approach to Mechanism Design

At present, major social media platforms (e.g. Facebook, Twitter, Instagram) do not have precisely-defined objectives. Instead they have a variety of mechanisms – content moderation teams, high-level oversight boards, hundreds of teams overseeing distinct sub-metrics – that are designed to maximize high-level KPIs or provide monitoring of system outputs. Beyond very high level performance metrics ("maximize engagement"), there is little understanding of or consensus on what a desirable recommender system for millions of people would look or feel like. While lacking a clear definition of a "good" system poses problems, it does not impede more proactive approaches. In particular, we need policies that will prepare platforms for a future specification of objectives. We outline two of these below: (1) focusing on pro-social metrics; (2) implementing principles of procedural fairness to platform governance.

With respect to metrics, we are mindful that there may be strong disagreements between users about how and where optimization constraints should be set. In a multiethnic society, let alone a global interface for billions of people, no single group has a monopoly on the definition of legitimate constraints on any part of the social fabric. That there will be disagreement on those metrics is not a flaw: it is only by engaging with those hard questions that meaningful interventions can be built into iterative technology systems.[158] The problem faced by platforms today is how to distinguish productive (pro-social) and unproductive (declining social trust, alienation) forms of disagreement within the actual functioning of recommender systems. We are encouraged by recent work on metrics for depolarization[159] and community well-being[160] that argue that this problem is tractable. The difference these metrics make is the ability to measure how users relate to each other at scale, rather than try to evaluate their subjective experience or satisfaction in a social vacuum. Rather than implicitly elevate the satisfaction of one group over others, it is possible to recommend content across populations in ways that demonstrably do not come at each other's expense.

With respect to governance, integrating principles of procedural fairness into content recommendation channels would at least lay the groundwork for evaluating cross-population outcomes. Such a process necessitates prior establishment of standards across products, consistent public reporting of ultimate findings of cross-platform tests (with privacy protections and provisions), and reasonable due process for any specific issue or implementation. We expect procedural fairness would also improve organizational integrity within tech firms, and direct feedback from former employees of companies like Meta and TikTok underline this perspective. Most technologists are justifiably proud of their contributions to the products and services they contributed to, including trust and safety and policy contributors who often see themselves as oppositional to product leadership. Without better leverage to influence design goals – and the compensation incentives of those who spend their professional lives pursuing those goals – it is hard to make progress.

---

[158] For more on the explicitly political dimensions of this question, see Dobbe, Roel, Thomas Krendl Gilbert, and Yonatan Mintz. "Hard choices in artificial intelligence." *Artificial Intelligence* 300 (2021): 103555.

[159] Stray, Jonathan. "Designing recommender systems to depolarize." *First Monday*, Volume 27, Number 5 - 2 May 2022. https://dx.doi.org/10.5210/fm.v27i5.12604

[160] Stray, J. Aligning AI Optimization to Community Well-Being. *Int. Journal of Com. WB* 3, 443–463 (2020). https://doi.org/10.1007/s42413-020-00086-3

Finally, we are encouraged by the recent prominence of generative AI models that attempt to specify desired performance in an objectives-first manner. Anthropic's Claude chatbot is an interesting contrast to OpenAI's ChatGPT, as the former presents itself as explicitly optimizing for "helpfulness" and "harmlessness" in its responses to user queries, rather than simply minimizing a static cost function within an arbitrarily large model.[161] These developments, and related open avenues for research, suggest both that current refinements to today's platforms are implementable, and that the next generation of recommender systems may well be able to integrate a more precise specification of objectives at the platform level. We would still need to see external, validated assessments of the claims of platforms to confirm the implementation of suitable accountability infrastructure – steps that generative AI platforms must take in order to demonstrably align with our proposal.

## Power Dynamics & Individual Autonomy

Our proposal draws inspiration from critiques of online platforms that have long been a focus of activists and scholars in the fields of media studies, human-computer interaction, and AI ethics. In particular, these researchers have highlighted how the unique affordances of platforms must be taken seriously, even as their effects on people remain poorly understood and in need of sustained investigation.

In these fields, platforms are often understood as *polysemic*, i.e. they support a finite but large range of interpretations of posted content,[162] and there is justified skepticism in measuring the direct effect of posted content on users.[163] Through this lens, content moderation is often portrayed as a critical – though limited – tool that is at present inconsistently used, misused, and

---

undervalued by major tech platforms.[164] These intuitions have led to key insights that lie at the basis of our framework:

- behavioral metrics alone are insufficient for assessment of harm prevention;
- the status quo unfairly places the burden on individuals to address issues that should be handled through broader institutional mechanisms;
- different subgroups may have incommensurable notions of what is good or harmful, and this needs to be reflected in any valid conception of design;
- platform infrastructure must be *experiment* and *intervention-driven*, not just data-driven.

In general, we are sympathetic to the need for qualitative evaluation, interpretive methods, analysis of power dynamics, and normativity, but feel they are better approached through regulation. Put more directly, we see these issues as a consequence of *platform design choices* that could be made differently but are presently not interrogated. Accountability infrastructure is needed so that the social dynamics wrought by platforms can even be represented and causally understood. In this way, a coherent and effective regulatory procedure could be followed and techno-solutionism could be rejected on both scientific and political grounds. In sum: we believe that if properly calibrated, the kinds of regulatory regimes described here would functionally shift power towards those who have been marginalized, in the form of direct assessments and protections for those communities. If implemented, these systems would likely induce leadership organizations, for instance, to prioritize community-driven design practices much earlier in the process.

Still, certain assumptions behind these critiques are in conflict with our proposed framework. In particular, some scholars operate within disciplinary frames that prioritize further investment and staffing in content moderation in lieu of more technical approaches to preventing platform harms at scale.[165] Some scholars also insist on power dynamics as the primary coherent organizing frame for platform oversight.[166] Moreover, harms are sometimes interpreted as content-specific and historical, predating the existence of platforms in ways that cannot be approached using technology itself.[167] Together, these assumptions have led scholars to avoid or dismiss normative questions about how platforms should or should not work instead of interrogating how companies could build platforms in ways that structure these dynamics differently.[168]

---

[164] Marwick, Alice. "Why do people share fake news? A sociotechnical model of media effects." *Georgetown Law Technology Review,* 2 474 (2018). https://georgetownlawtechreview.org/wp-content/uploads/2018/07/2.2-Marwick-pp-474-512.pdf
[165] Ibid.
[166] Anderson, CW. "Fake News is Not a Virus: On Platforms and Their Effects." *Communication Theory*, Volume 31, Issue 1, February 2021, pp. 42–61, https://doi.org/10.1093/ct/qtaa008
[167] Lenoir, T. and C Anderson. (2023). Introduction Essay: What Comes After Disinformation Studies. *Center for Information, Technology, & Public Life (CITAP), University of North Carolina at Chapel Hill*. Retrieved from https://citap.pubpub.org/pub/oijfl3sv
[168] Docherty, Neil and Asia Biega. "(Re)Politicizing Digital Well-Being: Beyond User Engagements." CHI '22, April 29-May 5, 2022, New Orleans, LA, USA. https://dl.acm.org/doi/pdf/10.1145/3491102.3501857

## Mechanism Design vs. Interface Design

Recourse to AI is sometimes offered in response to the problems discussed in this paper. Mark Zuckerberg himself named it as a solution to disinformation in his congressional hearings,[169] and Meta has continued to develop tools in this direction.[170] The argument goes that human-led content moderation is limited (there either are not enough humans, or the risks to humans are too high, or they are not able to be paid, or the harms may occur too quickly), so AI should instead be trained to handle content harms at scale.

While provocative and useful for the largest platforms (especially for acute cases like terrorism or copyright), this approach reduces structural harms to an optimization problem that can be solved by a highly-accurate classifier. But present computational and organizational constraints reveal the limits of this strategy: the largest language models now contain hundreds of billions of parameters,[171] while the latest training methods, including reinforcement learning from human feedback, require tens of thousands of human-generated labels for even low-end models.[172] Automated methods will still require some form of human oversight for data labeling, and recent roadblocks in the development of large language models[173] indicate the human infrastructure for such a project may not yet exist.

The mechanism design approach we favor does not automatically account for this complexity. But unlike other approaches that focus on content-level evaluation (either human- or AI-led), it does favor experiments and interventions that are sensitive to the effects of interface design. Platforms could test how the prevalence of structural harms are impacted by different interface designs or content limits (such as the recently announced "limits" for children[174]). More broadly, they could observe how different interfaces affect the type and spread of content amongst user populations regardless of harm. The merit of this approach is that it makes the complexity of platform oversight tractable by controlling for particular mechanisms and observing the result.

---

# Speech and Legal Considerations

This paper is not a legal analysis, and our purpose is not to situate specific claims in the context of constitutional arguments, or current reconsiderations of the Telecommunications Act. More so, we have three immediate rejoinders to First Amendment critiques of this approach prior to engaging on the merits: first, as discussed repeatedly, we believe these methods could be incorporated without government intervention via product leaders directly. Second, this approach could be taken up by foreign jurisdictions with different regulatory requirements regarding speech, especially the European Union under the DSA and the UK under its Online Safety bill. Third, we believe much of the motivation of this structure could be accomplished domestically in the US strictly through reporting requirements (akin to safety rules or anti trafficking rules which require independent auditing but not specific decision-making) – on the assumption that clear public evidence of structural harms would be sufficient to induce platforms to update procedures even without direct statutes.

Nonetheless, a common question raised at the suggestion of architecture features such as recommender systems relate to First Amendment limitations, on the grounds that limits on these systems amount to rules guiding editorial decisions of material protected by free speech restrictions.[175] We understand this concern: indeed, it was one of our motivations for initially focusing on architecture over content. We believe it is imperative to develop leverage points to address harms; without infrastructure of the type we propose, the main options for harm mitigation are facial textual analysis of content in the form of moderation (whether for removal or for reduction in distribution). Direct evaluation of content moderation *definitionally requires* an acceptance or rejection of content based on the speech itself, even within a "neutral" adjudication strategy favored by the Supreme Court.[176]

If executed properly, by contrast, the architectural approach we advance here never engages with any particular speech choice made by any individual, and instead only addresses aggregate population effects from technology choices which operate on that speech. Importantly, no speaker is prevented from expressing their viewpoint under this regime (outside of existing moderation procedures/standards); and there is no direct limitation even in a distributional sense for any viewpoint outside of an unrelated, principled process that is developed through a predefined, democratic decision-making process. For example, a legislative decision to limit exposure to material that reduces mental health of children might initiate this type of procedure. In this way, the regulatory process is organized to only respond to a health issue (mental health in this case) and not any specific facial analysis of content.

---

[175] Balbuzanova, Veronica. *First Amendment Considerations in the Federal Regulation of Social Media Networks' Algorithmic Speech, Part I*, AMERICAN BAR ASSOCIATION (January 29, 2021)
[176] For example: "McCullen v. Coakley" 134 S. Ct. 2518 (2014).
http://harvardlawreview.org/wp-content/uploads/2014/10/mccullen_v_coakley.pdf

Absent new Congressional actions, one might ask how legacy standards such as Section 230 of the Telecommunications Act relate to this, particularly in light of current litigation stemming from the Florida and Texas laws.[177] It is, perhaps, important to reiterate that the regime described here would not obviate the need for moderation of acute harms – and so empowering platforms to vigorously defend users from exposure to the most objectionable material, like CSAM or inducements of violence, will still remain critical. For those reasons, Section 230 protections enabling platforms to monitor their systems still would remain necessary. We do believe, though, that the kinds of framework outlined here could establish requirements imposed on platforms (especially for reporting) as part of the protections they are afforded by Section 230.

In other words, it is important for platforms to have protections enabling them to take appropriate steps to moderate content; but because moderation is insufficient to address structural harms, Congress could incorporate requirements for population assessments into the same regime as part of its protections of platforms.

## First Amendment Responses

Many scholars are currently pursuing and evaluating First Amendment objections to various proposals to limit harms caused by platforms.[178] We have solicited considerable legal discussion and analysis in the course of this work, and an interesting thread in First Amendment jurisprudence has come through those discussions. Specifically, if the platforms – and courts – regard selecting the parameterization of algorithms as free speech (by platforms),[179] then those algorithms definitionally are not protected by Section 230 (which instead governs third party speech). We would then be posed with a question of what reasonable limits could be placed on that type of speech.

We do not propose a specific scrutiny standard that would be appropriate for this question. Rather, by design, our infrastructural approach would be binding only when *demonstrable societal harms* are found.[180] Furthermore, under this system, accountability mechanisms would be implemented in the most targeted and least intrusive manner possible. Namely, it would allow companies to use existing infrastructure, procedures, and optimizations, and only be triggered in the event of a demonstrated harm (as "narrowly tailored" as possible an intervention).

---

[177] For example: "Free Speech Challenges to Florida and Texas Social Media Laws." *Congressional Research Service, Legal Sidebar*. 22 Sept. 2022. https://crsreports.congress.gov/product/pdf/LSB/LSB10748

[178] See for example Volokh, Eugene. "Treating social media platforms like common carriers?." *J. Free Speech L.* 1 (2021): 377.

[179] Balbuzanova, Veronica. *First Amendment Considerations in the Federal Regulation of Social Media Networks' Algorithmic Speech, Part I*, AMERICAN BAR ASSOCIATION (January 29, 2021).

[180] This would definitionally be a "compelling state interest" enacted by Congress – for example, protecting the mental health of millions of U.S. citizens.

Whether the Court would be persuaded by this structure surely is an open question; the intent of this proposal, though, is manifestly to approach the challenge of structural harms in a manner consistent with the passage of a scrutiny test.

### News & Media

Another question that might be asked is the extent to which this type of procedure would apply to news organizations. A first consideration set would be the size parameter outlined in Section 3, which we doubt any current news organization would be subject to. But assume, for a moment, that an organization like the *New York Times* was subjected to oversight.[181] The company maintains an independent editorial structure, inclusive of reporters and editors; it also maintains a business unit responsible for advertising, website maintenance, and distribution inclusive of algorithmic promotion choices. No accountability regime would be able to limit or prevent the Times from writing whatever stories it might choose. But it might apply to certain structural features of the app, such as the division at the Times that decides which articles to show at the top (including A/B testing), if such decisions were shown to have a measurable impact on user populations.

That this might be the case should not be surprising: choices made by business units already are subject to regulation, and business units already can trigger regulation in numerous ways via promotional choices. Examples of these types of choices might include promotion of fraudulent ads interspersed among legitimate articles written by the newsroom, or algorithmic promotion using illegally-sourced personal data of users, or employing a race-based segmentation to differentiate which users saw relevant information about housing availability. This is all to say: we already allow normal limitations on business choices of content producers, including news sources, so long as those limitations relate to non-speech based harms. In the event that the *New York Times* choice of algorithmic promotion of its independent editorial production did somehow fall afoul of the kinds of institutional harms outlined here, we believe it would be subject to the same rules and regulations.

## Self Governance, Policy Proposals, and Potential Rights

Present debates have focused on whether a government regulator must step in and fix what is wrong with how platforms are set up to operate, or if tech firms must simply do a better job policing the content they allow to propagate on their own platforms. We believe this debate is important. However, we also feel a more basic focus on platform access and tools for diagnostics is prerequisite to normative deliberations about what is or is not legitimate for platforms to do. Below we address some outstanding features of this problem.

---

[181] We doubt the New York Times would be, even if it was large enough–standards of journalistic integrity and critical feedback from readers already reflect some key dimensions of accountability infrastructure.

## Organizational Accountability

In the framework we have outlined, data collection would be tied to specific proposed harm assessments. This means that data will be oriented around defined experimental setups primarily, rather than general data under management by a platform. While broadening access to data is a critical step, it will be the association with product decisions that enables direct assessments associated with actionable outcomes, rather than pure (though potentially very important) research questions.

To achieve organizational accountability, the major outstanding question is how this association must be structured in order for interventions to be held accountable in an experimental sense; in other words, how the consequences of interventions can be reliably traced back to design choices. A few different technical interfaces are available to proceed.

One is the infrastructure for AI documentation underway at Hugging Face, including a major push for Model Cards as a baseline form of accountability for the use and deployment of trained machine learning models.[182] Another is Reward Reports, a form of AI documentation that regularly updates how the system specification (including optimization criteria) interacts with system behaviors over time.[183] This approach potentially connects specific design decisions – such as the addition or refinement of metrics as target parameters – to impacts on human populations. Model Cards and Reward Reports can be thought of as tools to support platform accountability, making either system components more transparent to third parties or the system as a whole more transparent to those who develop and manage it.

For Reward Reports to directly support organizational accountability, they would have to do more than passively record how a given optimization technique operates. They would have to include a log of who was responsible for particular implementations, why those techniques were chosen, and what effects would have to be observed for those implementations to be rolled back or modified. This is why fully-documented Reward Reports would include sections for:
- Optimization Intent (which catalogs performance metrics and failure modes),
- Institutional Interface (which asks which external entities the system will engage with and how the system will remain accountable to these stakeholders),
- Implementation and Evaluation (disclosing design elements critical to how data is processed and how user feedback is translated into performance metrics and fine-tune the system)
- System Maintenance (containing a changelog where designers log any changes made to the ML system over time)

---

[182] Mitchell, Margaret, et al. "Model cards for model reporting." *Proceedings of the conference on fairness, accountability, and transparency*. 2019.
[183] Gilbert, Thomas Krendl, et al. "Reward reports for reinforcement learning." *arXiv preprint arXiv:2204.10817* (2022).

## Establishing Rights

The nature of a mechanism design argument is that it is procedural. In a world of significant uncertainty – the existence of discrete harms caused by social media platforms remains highly contested[184] – we see that as an asset. We believe, normatively, that in a free society the legitimacy to conduct interventions should be conditioned upon the realization of socially relevant effects on specific populations. And yet, in other historical contexts, these types of antecedent mechanisms have resulted in specific consensus around protections for those populations which map onto other, less instrumental, frameworks. For instance, if it could be established that certain product features are overwhelmingly likely to result in harms to teen mental health, one could reframe a prior restraint provision against such a product feature as a protection of the rights of children.

Perhaps most concretely in context of the argument we have put forward, access to clean water has been established by the United Nations as a human right.[185] Countless people globally still live without this right – including in developed countries.[186] Domestically, the success of the Clean Water Act has transformed a bureaucratic assessment of policy standards into a generalized expectation of public services – quite a shift from the miasmists of the nineteenth century! This type of shift does not undercut the mechanistic assessments still needed for understanding what defines acceptably safe water; instead, it constitutes an organizing principle for civic participants seeking to ensure that access.

Such a shift might mirror transitions in certain other regulated industries. Consider pharmaceutical drugs, financial securities, or even something as simple as pedestrian foot traffic in a city. For drugs regulated under the FDA, there is no "right" of generalized access for care, nor established expectation that all drugs will be effective. Over time, however, there has been a functional codification of the duty of care rights that doctors are ethically bound to under the hippocratic oath: namely, potential users of drugs might express belief in having a right that taking a prescribed medicine is genuinely in the individualized best interest of the patient (outside of cases like drug trials which require specific patient knowledge and approval). For financial securities, rules tied to accredited investing similarly are situated as establishing baseline rights for citizens in society – namely that all citizens have baseline protections from excessive risk they are not positioned to take on, even if well intentioned. In transportation, public infrastructure access is often in tension with private interests; those intangible conflicts may or may not take the form of cost-benefit analysis. Over time, the establishment of standardized procedures induces a belief in baseline expectations which are often mapped onto rights frameworks (a right to equal access to job opportunities, for instance, in a city like New York).

## Recent Motivating Examples

It might be argued that our proposal for accountability infrastructure is not applicable to the types of problems faced by societal-scale platforms. In particular, perhaps either the conceptual lessons from past historical periods may not apply to the present moment, or the empirical basis for our proposal – that the impacts of algorithms on populations are directly measurable – would be so prohibitively difficult to demonstrate that it is unworthy of pursuit. We address each worry below.

Outside of prior scholarship, one of the authors conducted a series of experiments which informed motivations for this paper; we provide a couple examples of that work below. These examples were sourced from randomized trials in controlled laboratory environments under which "treatment" populations were exposed to discrete pieces of digital content (generally identical to content which received distribution on social media platforms), and comparing those results to responses by control groups who were exposed to an unrelated "neutral" piece of content.

The goal of these experiments was never to approximate the exact mechanisms which exist on platforms, or real-world effect sizes. Doing that would require the infrastructure described in this paper. Instead, these experiments were intended to assess an upstream question: to what extent can direct evidence be developed that exposure to content produces effects suitable for measurement of the type we propose? The nature of these experiments is such that it is impossible to answer these questions conclusively, but they do give reasonably strong directional evidence.

We include below two simple examples from Fellow Americans and the Fellow Americans Education Fund, non-profit entities which create and assess the persuasion effects of digital content.

### COVID-19 PSAs

The evolution of the COVID-19 pandemic has illustrated the connection between effective interventions and causal infrastructure all too concretely in recent years. A hallmark of the US response to the pandemic has been the spread of misinformation, conspiracies, and distrust of public officials nominally responsible for setting policy to protect the lives of citizens.[187] Different US states pursued rival policies to halt or allow the spread of COVID, and senior elected officials affirmed or denied its very existence.[188] In both scale and intensity, the harm of this contagion is as much cause for concern as the virus itself.

Still, what has not changed is the importance of metrics. Daily infection rates at the national, state, and local level have been vital to public understanding the virus over time. Vaccines have been enormously effective, not just because they were widely available but because their effectiveness was rigorously tested and widely known to critical institutional agencies and stakeholders. Even as national rates became less reliable with the rise of at-home testing, sewage contaminants have become a critical vector for measuring the spread and emergence of new COVID variants in close to real time.[189] In many ways, the politics of COVID have been least problematic where metrics were created to impartially evaluate the effectiveness of available interventions. That this is true for a virus that within weeks went from non-existent to global contaminant is a basis for hope that a similarly adaptive harm mitigation strategy for the affordances of societal-scale platforms is possible.

In late 2020 into early 2021, the group sought to produce and test content designed to induce the American public to take the warnings of public health officials seriously and follow CDC guidelines, culminating in getting vaccinated. Many of these messages (based on the randomized controlled experiments) were effective, shifting statistically significant shares of the population (compared with the control groups) to understand the importance of public interventions. These results were then incorporated in public distribution of content. An interesting trend emerged in some of these experiments, however: the most effective messages involved first-person accounts by health professionals, especially nurses, speaking directly to the camera about the lived experience of hospital emergency rooms during the early phase of the pandemic – in particular, about how those experiences were worsened by individuals' decisions to disobey safety rules.

While effective at shifting views, these messages also had an unintended effect: they tended to *reduce* measures of social trust – perhaps unsurprising, given that these sorts of PSAs are literally stories about civic peers not doing the right thing in a manner that endangers the lives of hospital staff. Later, subsequent messages were tested which were still effective at advocating

---

policy intentions without diminishing trust in the same way, and were recommended. These messages tended to lift up success stories, centering on the *agency* of participants to improve outcomes and celebrate those successes, building on strategies in the literature associated with group effects; these messages tended not to highlight failings (either institutional or individual) in the same manner.

These experiments illustrated two phenomena. First, exposure to these sorts of PSAs had measurable effects at least in immediate controlled settings, albeit with strong expectations of time-based dropoff effects. Second, they showed that even for explicitly pro-social and well-intentioned content, it was possible to induce deleterious effects on a social measure (trust). In this case, such a tradeoff may well have been worth it – we believe it was – in the form of stronger advocacy to support an important public health response. But in online settings, such a countervailing argument would not be so readily apparent.

## 2021 Superbowl Commercials

In a second example, in February of 2021, the Fellow Americans team (in collaboration with the content assessment company Swayable) tested a slate of 25 Superbowl commercials put out directly for the game. These messages were apolitical and were explicitly designed to foster brand affinity; but given the timing of the game and the pandemic, the spots were also quite clearly intended to be contextual to the COVID pandemic.

The analysis showed a few key findings. There was an unexpectedly strong connection between perceptions of the social effects of brand and purchase intent. In this study, there was a direct positive correlation of .87 between these two metrics – suggesting a very strong connection (whatever the causal reason) between social value and sales.[190] When looking at measures of trust, there was a much higher variance in outcomes – with some content boosting trust and others indicating negative effects. The dataset was too small to reach broad-reaching conclusions about systematic outcomes, but there were some striking indications embedded. Specifically, two takeaways were indicated. First, the attributes of content that tended to move social trust more positively in a corporate context mirrored the content attributes in other areas (like the COVID example just described, or in politics). Positive tones, optimistic descriptions of struggle, and first person accounts from ordinary Americans telling their stories each tended to move trust metrics more positively. And second, exposure to this type of content tended to neutral to small negative metrics on some measures of institutional trust, an unexpected finding and one that would need to be replicated further. But it suggested that repeated exposure to these types of content can prime viewers in their perceptions of the roles and obligations of members of society.

---

[190] Without a doubt, this effect would be smaller now outside of the COVID context of early 2021.

*As the tide of chemicals born of the Industrial Age has arisen to engulf the environment, a drastic change has come about in the nature of the most serious public health problems. Only yesterday mankind lived in fear of the scourges of smallpox, cholera, and plague that once swept nations before them. Now our major concern is no longer with the disease organisms that once were omnipresent; sanitation, better living conditions, and new drugs have given us a high degree of control over infectious disease. Today we are concerned with a different kind of hazard that lurks in our environment — a hazard we ourselves have introduced into our world as our modern way of life has evolved.*

<div align="right">

– Rachel Carson, *Silent Spring*[191]

</div>

# CONCLUSION

We have claimed that an evidence-based approach to evaluating societal-scale platforms is implementable and effective at mitigating structural harms. We do not assert to have "solved" the deeper normative problems associated with this regime, even as we have suggested that opportunities to mitigate harms associated with mental health and societal trust offer immediate pathways forward. However, regardless of specific metrics, accountability infrastructure of this kind is necessary for aligning available interventions with knowledge of their consequences among affected populations. Absent that knowledge, public policy will remain in critical ways a product of guesswork about how platforms actually work, let alone how they impact their users.

The key lesson of public health is that societal-scale problems can be made tractable through new infrastructural commitments. While new laws in a diverse range of countries have sought to tackle different challenges posed by technology platforms, the history of prior epochs suggests that these efforts will be mistargeted unless they are scoped to directly evaluate the harms they seek to mitigate. Even if the bureaucratic needs of this infrastructure would be significant, most current platforms are already well-suited to the technical implementation of our proposal.

Updating institutions to match the need of the most recent advances in technology is never an easy task. The framework described here would ideally involve (in an American context) collaborations between Congress and an executive agency like the FTC, independent researchers and universities, and technologists and product managers/designers within large companies. It would require reaching consensus on the collective problems we believe constitute a societal interest worthy of placing limits on industry – beginning with the health of children – and settling on evaluation schema to chip away at the worst abuses.

And yet with thousands of different people already working on interdisciplinary approaches to the evaluation and response to these technology platforms, there is already a growing awareness and desire for real interventions – especially those that avoid, as much as possible, direct restrictions on free expression. As a community, we can heed the caution sounded by *Silent Spring* sixty years ago and recognize that modern life has seeded new hazards that we did not anticipate – and take action to decisively reduce their worst effects.

---

[191] Carson, Rachel. *Silent Spring*. Penguin, 2002. p. 187

# ACKNOWLEDGEMENTS

The authors thank the Berkman Klein Center at Harvard University and the Digital Life Initiative at Cornell Tech for serving as supportive homes for this project. They also thank the *MIT Technology Review, The Atlantic*, and the 2022 ACM Conference on Equity and Access in Algorithms, Mechanisms, and Optimization (EAAMO) for providing feedback and for publishing versions of this work.

The authors would also like to thank John Basl, James Grimmelmann, Emanuel Moss, Nathan Freitas, Claire Stapleton, Kate Klonick, Jonathan Zittrain, James Mickens, Rebecca Rinkevich, Mateus Guzzo, Helen Nissenbaum, Ravi Iyer, Danielle Whelton, and Daniel Adler, for their feedback and contributions, and Jillian Maryonovich for design support.

# KEY REFERENCES